\documentclass[aps,prb,superscriptaddress,showpacs,twocolumn]{revtex4}
\usepackage{epsfig,graphicx,times}
\newcommand{\Uext}{U_{\rm ext}}
\newcommand{\Utf}{U_{\rm TF}}
\newcommand{\Htf}{\hat H_{\rm TF}}
\newcommand{\Uks}{U_{\rm KS}}

\newcommand{\Vcoul}{V_{\rm coul}}
\newcommand{\Vxc}{V_{\rm xc}}
\newcommand{\Vsc}{V_{\rm screened}}
\newcommand{\Vbare}{V_{\rm bare}}

\newcommand{\Etf}{E_{\rm TF}}
\newcommand{\Est}{E_{\rm ST}}
\newcommand{\Eststar}{E_{\rm ST^*}}
\newcommand{\Eks}{E_{\rm KS}}
\newcommand{\Eop}{{\cal E}_{\rm 1p}}

\newcommand{\Eosc}{{\cal E}^{\rm osc}_{\rm 1p}}

\newcommand{\Fks}{{\cal F}_{\rm KS}}
\newcommand{\Ftf} {{\cal F}_{\rm TF}}
\newcommand{\Ttf}{{\cal T}_{\rm TF}}
\newcommand{\Tks}{{\cal T}_{\rm KS}}

\newcommand{\Exc}{{\cal E}_{\rm xc}}
\newcommand{\Etot}{{\cal E}_{\rm tot}}

\newcommand{\nKS}{n_{\rm KS}}

\newcommand{\nKSosc}{(n_{\rm KS})_{\rm osc}}
\newcommand{\nTF}{n_{\rm TF}}
\newcommand{\tn}{\tilde n}
\newcommand{\tnosc}{{\tilde n}_{\rm osc}}

\newcommand{\muTF}{\mu_{\rm TF}}

\newcommand{\exc}{\epsilon_{\rm  xc}}
\newcommand{\va}{v_a}
\newcommand{\vb}{v_b}

\newcommand{\kf}{k_F}

\newcommand{\br}{{\bf r}}
\newcommand{\bR}{{\bf R}}
\newcommand{\bl}{{\bf l}}

\newcommand{\bp}{{\bf p}}
\newcommand{\bq}{{\bf q}}

\newcommand{\al}{\alpha}
\newcommand{\be}{\beta}
\newcommand{\s}{\sigma}
\newcommand{\e}{\epsilon}
\newcommand{\et}{\tilde \epsilon}

\def\lsim{\hbox{\lower .8ex\hbox{$\, \buildrel < \over \sim\,$}}}
\def\gsim{\hbox{\lower .8ex\hbox{$\, \buildrel > \over \sim\,$}}}

\begin{document}

\title{ 
Landau Fermi Liquid Picture of Spin Density Functional Theory: \\
Strutinsky Approach to Quantum Dots
}

\author{Denis Ullmo}
\affiliation{Department of Physics, Duke University, Durham, North
  Carolina 27708-0305}
\author{Hong Jiang}
\affiliation{Department of Physics, Duke University, Durham, North
  Carolina 27708-0305}
\affiliation{Department of Chemistry, Duke University, Durham, North
  Carolina 27708-0354}
\author{Weitao Yang}
\affiliation{Department of Chemistry, Duke University, Durham, North
  Carolina 27708-0354}
\author{Harold U.~Baranger}
\affiliation{Department of Physics, Duke University, Durham, North
  Carolina 27708-0305}

\date{January 20, 2004}

\begin{abstract}
We analyze the ground state energy and spin of quantum dots obtained
from spin density functional theory (SDFT) calculations.  First, we
introduce a Strutinsky-type approximation, in which quantum
interference is treated as a correction to a smooth Thomas-Fermi
description.  For large irregular dots, we find that the second-order
Strutinsky expressions have an accuracy of about 5 percent compared to
the full SDFT and capture all the qualitative features.  Second, we
perform a random matrix theory/random plane wave analysis of the
Strutinsky SDFT expressions.  The results are statistically similar to
the SDFT quantum dot statistics.  Finally, we note that the
second-order Strutinsky approximation provides, in essence, a Landau
Fermi liquid picture of spin density functional theory.  For instance,
the leading term in the spin channel is simply the familiar exchange
constant.  A direct comparison between SDFT and the perturbation
theory derived ``universal Hamiltonian'' is thus made possible.
\end{abstract}



\maketitle

\section{Introduction}

Semiconductor quantum dots are now routinely obtained using
electrostatic gates or etching processes to pattern a two dimensional
electron gas formed in some heterostructure (typically
GaAs/AlGaAs).\cite{Kouwenhoven97} Many of their sometimes surprising
properties are now reasonably well understood from a qualitative or
statistical point of view,\cite{Kouwenhoven97,Alhassid00RMP} and it is
now realistic to think about using these quantum dots for some
specific purpose, such as spin filtering,\cite{Weinmann95PRL} current
or spin pumping,\cite{Mucciolo02PRL} or in the context of quantum
information.\cite{loss98}

In this context, it becomes important to go beyond a qualitative or
statistical description, and to develop tools able to predict
quantitatively the properties of a specific quantum dot for given
parameters.  For isolated or weakly connected dots -- the Coulomb
Blockade regime -- the Coulomb interaction between electrons plays an
important role and has to be taken into account properly.

A method of choice in this regard is the density functional
approach,\cite{ParrYang,Jones89RMP} which has been widely used in
theoretical modeling of quantum dots. \cite{Reimann02RMP} More
specifically, since it is necessary to describe correctly the spin
degree of freedom, we have in mind a spin density functional, where
each density of spin $n^\s(\br)$, with $\s = \alpha, \beta$
corresponding to majority and minority spins, are treated as an
independent variable. How this can be achieved in practice for a dot
containing up to four hundred electrons, and for an $r_s$ parameter as
high as four, has been demonstrated in a series of
papers.\cite{Jiang03,Jiang03b,Jiang04}

One striking feature of these calculations, however, is that the
qualitative picture which emerges is somewhat unexpected in view of
previous results.  Indeed, within a statistical approach and assuming
the classical dynamics within the nanostructure is sufficiently
chaotic, one can model the wave functions in the quantum dot using
Random Matrix Theory (RMT).  If furthermore the Coulomb interaction is
treated within the Random Phase Approximation (RPA), it is possible to
derive various statistical quantities,\cite{prl_blanter_97,
prb_ullmo_01, prb_usaj_01,pr_aleiner_02} such as the distribution of
spacing between Coulomb Blockade conductance peak, or the probability
of occurrence of non-standard spins [that is, not zero (not one-half)
for even (odd) particle number].  It turns out, for instance, that the
spin density functional calculations give a larger number of ``high''
spins than was predicted within RMT plus RPA modeling.\cite{Jiang03}
Such discrepancies could originate from a variety of causes, ranging
from the the statistical behavior of the Kohn-Sham wave functions to
an intrinsic failure of one or the other approach.  The goal of this
paper is to clarify this issue.

For this purpose, we need a way to understand, or to organize, the
numbers obtained from the full-fledged spin density functional theory
(SDFT).  This can be done using the Strutinsky approximation to DFT
discussed in Ref.~\onlinecite{prb_tatsuro_01}, up to straightforward
modifications to include the spin variable.

The Strutinsky approach is an approximation scheme, developed in the
late sixties in the context of nuclear physics, in which the
interference (or shell) effects are introduced
perturbatively.\cite{np_strutinsky_68, rmp_brack_72} It has been since
then used in many subfield of physics,\cite{book_brack} including the
calculation of the binding energy of metal clusters.\cite{YannPRB93}
Standard treatments are usually limited to first-order corrections; in
this case, it is similar to the Harris functional \cite{harrisPRB85}
familiar to the DFT literature.  Here, in contrast, we use the
second-order approximation \cite{ElstnerPRB98} developed in
Ref.~\onlinecite{prb_tatsuro_01}. We shall see that the second-order
Strutinsky scheme turns out to be extremely accurate in some
circumstances.  Furthermore, even when it is less precise, which
happens in conjunction with the occurrence of ``spin contamination''
in the SDFT calculations, it still provides a qualitatively correct
statistical description.

The second-order Strutinsky corrections can be cast in a very natural
form:\cite{prb_tatsuro_01} in fact, they amount to taking into account
the residual (screened) interactions between quasiparticles in a
Landau Fermi liquid picture. They are thus amenable to treatment by
the same RMT approach as was used previously for RPA, providing an
analogue of the ``universal Hamiltonian'' in the case of SDFT. Since
within the Strutinsky approximation the quantum dot properties are
relatively transparent, we will then be in position to discuss the
difference between SDFT results and those obtained from RMT plus RPA.

The paper is organized as follows.  In Section~\ref{sec:strut}, we
briefly review the Strutinsky approximation as it applies to spin
density functional theory.  In Section~\ref{sec:T&C}, we consider more
specifically the local density approximation from this perspective,
and in particular the screened potential that it
implies. Section~\ref{sec:KSvsST} covers in detail the specific case
of a model quantum dot with quartic external confining potential.
This model is used to investigate the accuracy of the Strutinsky
approximation for various electron densities.  In
Section~\ref{sec:FL}, we discuss how the Fermi liquid picture
emerging from the Strutinsky approximation scheme
can be used, in conjunction with a random plane wave model of the
wavefunctions, to analyze the peak spacing and spin
the distributions resulting from the SDFT calculation.  This framework
also makes it possible to discuss the mechanism of spin
contamination.  Finally, in the last section we come back to the
original question motivating this work and make use of what we
understand from the Strutinsky approximation to discuss the
discrepancies between SDFT calculations and RPA plus RMT predictions.

\section{Strutinsky approximation for spin density functional
  theory}
\label{sec:strut}

In the spin density functional description of a quantum dot, one
considers a  functional of both spin densities
$[n^\alpha(\br),n^\beta(\br)]$
\begin{equation} \label{eq:Fks}
    \Fks[n^\al,n^\be]=\Tks [n^\al,n^\be]+\Etot[n^\al,n^\be] \; ,
\end{equation}
where $\s = \al,\be$ correspond to majority and minority spins,
respectively. In this expression, the second term is an explicit
functional of the densities,
\begin{eqnarray} \label{eq:def_Etot}
 \Etot[n^\al,n^\be]\equiv\int \!d\br \;n(\br) \;\Uext(\br) \qquad
\qquad \qquad \nonumber\\ + \int \! d\br d\br'
\,n(\br)\,v_\mathrm{int}(\br-\br')\,n(\br')+\Exc [n^\al,n^\be] \; ,
\end{eqnarray}
where $n(\br)=n^\al(\br) + n^\be(\br)$, $\Uext(\br)$ is the exterior
confining potential, and the precise form of the exchange correlation
term $\Exc [n^\al,n^\be]$ is to be discussed in more detail in
Section~\ref{sec:T&C}.  $v_\mathrm{int}(\br,\br')$ is the
electron-electron interaction kernel. The presence of metallic gates
can be taken into account by including an image term in addition to
the bare Coulomb interaction,
\begin{equation}
v_\mathrm{int}(\br-\br')=\frac{e^2}{|\br-\br'|}-\frac{e^2}{\sqrt{|\br-\br'|^2+4z_d^2}},
\end{equation}
where $z_d$ is the distance between the top confining gate and
two-dimensional electron gas. The image term in the interaction kernel
greatly reduces classical Coulomb repulsion between electrons at a
distance larger than $z_d$ so that the electron density far from the
boundary is quite uniform.  The bare Coulomb interaction can be
recovered by letting $z_d$ go to infinity.

The kinetic energy term $\Tks [n^\al,n^\be]$, on the other hand, is
expressed in terms of the  auxiliary set of orthonormal functions
$\psi_i^{(\al,\be)}$ ($i=1,\cdots,N_{\al,\be}$) such that
$ n^\s({\bf r}) \equiv \sum_{i=1}^{N_\s} |\psi^\s_i({\bf r})|^2$
as
\begin{eqnarray}
    \Tks[n^\al,n^\be] 
            &=&  {\hbar^2 \over 2m} \int
    \sum_{\s=\al,\be} \sum_{i=1}^{N_\s}
    \, |\nabla\psi^\s_i({\bf r})|^2 \, d{\bf r} \; .
    \label{eq:def_TKS}
\end{eqnarray}

>From the density functional Eq.~(\ref{eq:Fks}) , the ground state
energy of the quantum dot containing
$(N_\al,N_\be)$ particles of spin $(\al,\be)$ is obtained as
\begin{equation} \label{eq:Eks}
\Eks(N_\al,N_\be) = \Fks[\nKS^\al,\nKS^\be]
\end{equation}
 where   the  Kohn   Sham   densities  $[\nKS^\al(\br),\nKS^\be(\br)]$
minimize $\Fks$ under the constraint  given by the number of
particles of  each  spin.   This  in  practice is  equivalent  to
solving  the Kohn-Sham equations
\begin{equation}  \label{eq:KS}
    \left(-{\hbar^2 \over 2m} \nabla^2   +
    \Uks^\s({\bf r}) \right) \psi^\s_i({\bf r}) =
    \e^\s _i \psi^\s_i({\bf r}) \; , \quad
    i=1,...,N_\s \; .
\end{equation}
with the spin dependent self-consistent potential
\begin{equation} \label{eq:Uks}
  \Uks^\s (\br) = \frac{\delta \Etot}{\delta n^\s(\br)}
  [\nKS^\al,\nKS^\be]  \; .
\end{equation}

In this section, we shall give a brief description of the second-order
Strutinsky approximation as it applies to spin density functional
theory.  Up to the introduction of the spin indices, the derivation of
this approximation follows exactly the same lines as the spinless case
discussed in details in Ref. \onlinecite{prb_tatsuro_01}. We shall
therefore not reproduce it here, but rather try to emphasize what
exactly are the assumptions made in deriving the approximation, and
then merely write down the expression we shall use in the following
sections.

\subsection{Generalized Thomas-Fermi approximation}

The basic idea of the Strutinsky energy correction method is to
start from smooth approximations, $\Etf$ and  $\nTF(\br)$, to the
DFT energy $\Eks$ and electronic density $\nKS(\br)$.  Then,
fluctuating terms are added perturbatively as an
expansion in the small parameter
\begin{equation} \label{eq:def_delta_n}
  \delta n (\br) = \nKS(\br) - \nTF(\br) \; .
\end{equation}
In the original work of Strutinsky \cite{np_strutinsky_68,
rmp_brack_72} various ways of constructing the smooth approximation
have been considered.  The most natural choice for $\nTF^{\sigma}(r)$
turns out to be the solution of the Thomas-Fermi equation
\begin{equation} \label{eq:TF}
\frac{\delta \Ftf}{\delta n^{\sigma}} [n^\al,n^\be] = \muTF^{\sigma}
\end{equation}
coupled with $\Etf = \Ftf[\nTF^\al(r),\nTF^\be(r)]$.
The ``generalized'' Thomas-Fermi functional is defined as
\begin{equation}
  \Ftf [n^\al,n^\be] = \Ttf [n^\al,n^\be]+\Etot[n^\al,n^\be] \; ,
\end{equation}
with  $\Etot[n^\al,n^\be]$ given  by  Eq.~(\ref{eq:def_Etot}). It
differs from  the original  spin density functional  only in
that     the    quantum     mechanical    kinetic     energy    $\Tks$,
Eq.~(\ref{eq:def_TKS}), is  replaced by  an explicit functional  of the
density. For two dimensional systems this takes the form
$\Ttf[n^\al,n^\be]  = \Ttf^{(0)} [n^\al,n^\be] + \Ttf^{(1)}
    [n^\al,n^\be] $ with 
\begin{eqnarray}
\Ttf^{(0)} [n^\al,n^\be] & = & \frac{1}{2 N(0)} \int \! d \br
    \big[n^\al(\br)^2 + n^\be(\br)^2\big] \label{eq:TTF0} \\
\Ttf^{(1)} [n^\al,n^\be] & = & \frac{\eta}{8 \pi N(0)} \int \!\!d \br
    \left[\frac{(\nabla n^\al)^2}{n^\al} + \frac{(\nabla n^\be)^2}{n^\be}\right]
    \label{eq:TTF1}
\end{eqnarray}
where $N(0) = m/\pi\hbar^2$ the density of states for 2-dimensional
systems and $\eta$ a dimensionless constant taken here to be $0.25$.
Such a choice for the kinetic energy functional correctly takes into
account the Pauli exclusion principle, and thus that an increase in
kinetic energy is required to put many particle at the same space
location, but fails to include fluctuations associated with quantum
interference.

To  illustrate this, let  us   for a  short while  the assume
$\eta = 0$, i.e.\ only the first term $\Ttf^{(0)}$ of the
Thomas-Fermi kinetic energy is taken into account. Then, one can show
that the Thomas-Fermi density fulfills the self-consistent equation 
\begin{equation} \label{eq:TH_sc}
   \nTF^\s(\br) = \bar n^\s [\Utf^\s](\br)
\end{equation}
where the Thomas-Fermi potential is defined,  as in
Eq.~(\ref{eq:Uks}), by 
\begin{equation} \label{eq:Veff}
  \Utf^\s (\br) = \frac{\delta \Etot}{\delta n^\s(\br)}
  [\nTF^\al,\nTF^\be]
\end{equation}
and 
\begin{equation} \label{eq:weyl}
\bar n^\s [\Utf^\s](\br) =
\int \!\!\frac{d\bp}{(2\pi\hbar)^d}\,
     \Theta\left[ \mu^\s_{\rm TF} - \frac{\bp^2}{2m} - \Utf^\s(\br)\right]
\end{equation}
is the Weyl part of the density of states of a system of independent
 particles evolving under the potential $\Utf^\s(\br)$ ($\Theta$ is
 the Heaviside step function, $d=2$ is the dimensionality, and the
 chemical potentials $\mu^\s_{\rm TF}$ are chosen such as to fulfill
 the constraints on the total number of particles with spin $\al$ and
 $\be$). From its definition, $\bar n^\s [\Utf^\s](\br)$ (and as a
 consequence $\nTF^\s(\br)$) is a smooth function, in the sense that
 it can change appreciably only on the scale on which $\Uext(\br)$
 varies, but cannot account for the quantum fluctuations of the
 density on the scale of the Fermi wavelength.

The Weyl approximation is, however, only the leading term in a
semiclassical expansion of the smooth part of the density of states,
and higher-order corrections in $\hbar$ can be added in a systematic
way.  The standard way to choose $\Ttf^{(1)}$ \cite{book_brack} is
such that Eq.~(\ref{eq:TH_sc}) holds, but with an approximation to the
smooth density of states $\bar n^\s [\Utf^\s](\br)$ which includes
both the Weyl term Eq.~(\ref{eq:weyl}) and the first $\hbar$
corrections.  For two dimensional systems, however, the corrective
term computed from this prescription turns out to be zero, when the
presence of $\Ttf^{(1)}$ is actually useful in smoothing the
Thomas-Fermi density near the boundaries of the classically allowed
region.  We have therefore used the phenomenological Weis\"acker-like
term\cite{book_brack} Eq.~(\ref{eq:TTF1}) which plays a similar role.

\subsection{Strutinsky correction terms}

The practical implementation for the Strutinsky scheme can be
summarized as follows.  The first step consists in solving the
generalized Thomas-Fermi equation Eq.~(\ref{eq:TF}), which defines a
zeroth-order approximation for the ground state energy, $\Etf$, as
well as an approximation to the density of particles
$\nTF^{\sigma}(\br)$.  From this latter quantity, one derives for
each spin $\s = \al,\be$ the effective potential $\Utf^{\sigma} (\br)
$ through Eq.~(\ref{eq:Veff}).

The second step consists  in solving the Schr\"odinger equations (again
for each spin)
\begin{equation}  \label{eq:schroe}
    \left(-{\hbar^2 \over 2m} \nabla^2   +
    \Utf^\s({\bf r}) \right) \phi^\s_i({\bf r}) =
    \et^\s _i \phi^\s_i({\bf r}) \; , \quad
    i=1,\ldots,N_\s \, .
\end{equation}
Therefore, while the Kohn-Sham equations are both quantum mechanical in
nature and self-consistent, here all the self-consistency is left at
the ``classical-like'' level of the Thomas-Fermi equation, and the
quantum mechanical wave interference aspect is taken into account
without self-consistency.  One obtains in this
way a new density
\begin{equation} \label{eq:tilde_n}
  \tn^\s(\br) = \sum_1^{N_\s} | \phi^\s_i|^2 (\br) \; .
\end{equation}
It can be seen\cite{prb_tatsuro_01} that $\nTF^{\sigma}(\br)$ is {\em by construction} a
smooth approximation to $\tn^{\sigma}(\br)$ [this is basically the content of
Eq.~(\ref{eq:TH_sc})].  Therefore
\begin{equation} \label{eq:tilde_nosc}
  \tnosc^\s(\br) \equiv \tn^\s(\br) - \nTF^\s(\br)
\end{equation}
can  be considered  as  the  oscillating part  of  $\tn^\s$, and  will
describe  the short scale  variations of  the density  associated with
quantum interference.

Once the eigenfunctions and eigenvalues of Eq.~(\ref{eq:schroe})
are known, corrections to the Thomas-Fermi energy can be added
perturbatively
\begin{equation}
  \label{eq:strut_corr}
  \Eks \simeq \Etf + \Delta E^{(1)} + \Delta E^{(2)} \; .
\end{equation}
The first-order correction turns out to be simply the oscillating part of the
one particle energy\cite{prb_tatsuro_01,book_brack}
\begin{equation} \label{eq:DE1}
  \Delta E^{(1)} = \Eosc = \Eop - \bar \Eop \;
\end{equation}
with
\begin{equation}
  \Eop = \sum_{i = 1}^{N_\al} \et^\al_i + \sum_{i =
  1}^{N_\be} \et^\be_i
\end{equation}
the one particle energy, and
\begin{equation}
  \bar \Eop  = \Ttf[\nTF^\al,\nTF^\be] +
  \sum_{\s=\al,\be} \int \!d\br \,\Utf^\s(\br) \,\nTF^\s(\br)
\end{equation}
its smooth part.

The second-order correction $\Delta E^{(2)}$ can be expressed in
two different ways\cite{prb_tatsuro_01}
\begin{equation}
  \label{eq:DE2}
  \Delta E^{(2)} = \frac{1}{2} \!\sum_{\s,\s' = \al,\be}
  \int \!d\br d\br' \,\tnosc^\s(\br) \,\Vbare^{\s,\s'}(\br,\br')
   \,\delta n^{\s'}(\br')
\end{equation}
\begin{equation}
      \label{eq:DE2*}
   \Delta E^{(2)*} = \frac{1}{2}  \!\sum_{\s,\s' = \al,\be}
   \int \!d\br d\br' \,\tnosc^\s(\br)
   \,\Vsc^{\s,\s'}(\br,\br') \,\tnosc^{\s'}(\br')
\end{equation}
with
\begin{equation}
\label{eq:Vbare}
\Vbare^{\s,\s'}(\br,\br') =
  \frac{\delta^2 \Etot}{\delta n^\s(\br) \delta n^{\s'}(\br')}
       [\nTF^\al,\nTF^\be]
\end{equation}
\begin{widetext}
\begin{equation}
\label{eq:Vsc}
 \Vsc^{\s,\s'}(\br,\br') = \sum_{\s'' = \al,\be}
 \int d\br''
 \left[ \left( \frac{\delta^2 \Ttf}{\delta n^2} +  \Vbare
 \right)^{-1} \right]^{\s,\s''}
 \hspace{-15pt} (\br,\br'') \; \left[ \frac{\delta^2 \Ttf}{\delta n^2}
 \cdot\Vbare \right]^{\s'',\s'}\hspace{-15pt} (\br'',\br') \; .
\end{equation}
\end{widetext}
(The matrix inversion here should be taken with
respect to both the spatial position and the spin indices.)
Note that in Eq.~(\ref{eq:Vsc}) we  can use
\begin{equation}
  \frac{\delta^2 \Ttf^{(0)} }{\delta n^\s(\br) \delta n^{\s'}(\br') }
  =
  \frac{2}{N(0)} \delta_{\s,\s'} \, \delta(\br'-\br)
\end{equation}
and, neglecting terms involving derivatives of the Thomas-Fermi
density,
\begin{equation}
  \frac{\delta^2 \Ttf^{(1)} }{\delta n^\s(\br) \delta n^{\s'}(\br') }
  = -
  \frac{ \eta}{4 \pi N(0)} \frac{
  \delta_{\s,\s'}}{\nTF^\s(\br)}
  \, \Delta_{r'} \delta(\br'-\br)
  \; .
\end{equation}

The first of these second-order expressions, Eq.~(\ref{eq:DE2}),
involves the $\delta n^{\sigma}(\br)$ which are in principle unknown.
It is therefore not useful if one is trying to avoid performing the
full Kohn-Sham calculation.  Here, however, we shall also do this full
self-consistent calculation in order to evaluate the accuracy of the
Strutinsky scheme; we therefore know $\delta n^{\sigma}(\br)$.  In
this case Eq.~(\ref{eq:DE2}) turns out to be somewhat simpler to
compute than Eq.~(\ref{eq:DE2*}).  This is not, however, because of
the matrix inversion in Eq.~(\ref{eq:Vsc}) -- for the case we shall
consider, the screening length is much smaller than the size of the
system, allowing this inversion to be done analytically.  Rather, the
resulting screened potential will turn out to be diagonal neither in
the position nor in the momentum representation (see next section), in
contrast to $\Vbare^{\s,\s'}$ which is the sum of two parts, one
diagonal in the position and the other in the momentum representation.
Since Eq.~(\ref{eq:DE2}) and (\ref{eq:DE2*}) are essentially at the
same level of precision, we shall in the following mainly use the
first one Eq.~(\ref{eq:DE2}); we will apply Eq.~(\ref{eq:DE2*}) only
with an approximate $\Vsc$ which is diagonal in the momentum
representation, thus inducing additional errors that are not intrinsic
to the Strutinsky scheme.

\subsection{Applicability of second-order Strutinsky expressions}

Although we shall not reproduce here the derivation of
Eq.~(\ref{eq:DE2}) (see again Ref.~\onlinecite{prb_tatsuro_01} for
details), the condition of applicability of this equation can be
understood easily by comparing the Kohn-Sham equations
Eq.~(\ref{eq:KS}) with the ones defined by the Thomas-Fermi potential
$\Utf^{\sigma}(\br)$, Eq.~(\ref{eq:schroe}). We see there that the
Kohn-Sham wave functions $\psi_i^{\sigma}$ and their approximations
$\phi_i^{\sigma}$ are defined through Schr\"odinger equations that
differ only by a difference in the potential term
\begin{eqnarray} \label{eq:delta_U}
  \delta U^\s (\br) & = & \Uks^\s (\br) - \Utf^\s(\br) \\
  & \simeq & \sum_{\s'=\al,\be} \int d\br' \;\Vbare^{\s,\s'}(\br,\br') 
\;\delta
  n^{\s'}(\br') \nonumber
\end{eqnarray}
with $\Vbare(\br,\br')$ defined by Eq.~(\ref{eq:Vbare}).

The main approximation in the derivation of Eq.~(\ref{eq:DE2}) is that
the  $\delta U  (\br)$ can  be taken  into account  by a  second-order
perturbative  calculation.   In general,  this  implies  that the  non
diagonal  matrix elements  $ (\delta  U)^\s_{ij} =  \langle  \phi^\s_i |
\delta  U^\s |  \phi^\s_j \rangle$  ($i  \neq j)$  should typically  be
smaller  than  the mean  level   spacing  $\Delta$  between  one  particle
energies.   More specifically, since  the only
relevant errors  are ones modifying the  Slater determinant formed
by  the occupied  orbitals, the  actual condition  is that  the matrix
element  between the  first unoccupied  orbital and  the  last occupied
orbital is smaller than the spacing between these level, that is
\begin{equation} \label{eq:DE2_cond}
  \langle \phi^\s_{N_\s+1} | \delta U^\s | \phi^\s_{N_\s} \rangle
  \ll (\et_{N_\s+1} - \et_{N_\s}) \qquad (\s=\al,\be) \; .
\end{equation}
A good accuracy, on the scale  of the one particle mean level spacing,
of the Strutinsky approximation in the form Eq.~(\ref{eq:DE2}) is
equivalent to the above condition.

Now,  Eq.~(\ref{eq:DE2_cond})  again  involves $\delta
n^{\sigma} (\br)$, which is unknown.   It is therefore not possible to
prove  rigorously  that it  should  be  fulfilled. It is  possible,
however, to  show the consistency of such an  assumption. If  we assume  that
\begin{itemize}
\item[(i)] $|\langle \phi^\s_i | \delta U^\s | \phi^\s_j \rangle| \ll
\Delta$ for all pairs of orbitals $(i,j)$ and $\s=\al,\be$, so that
$\delta U^\s$ can be treated perturbatively, and
\item[(ii)] the oscillating parts of the Kohn-Sham densities
$\nKSosc^{\sigma}(\br)$ do not differ significantly from
$\tnosc^{\sigma}(\br)$, implying that $\delta n^{\sigma} (\br)
\simeq \delta \bar n^{\sigma} (\br) +\tnosc^{\sigma}(\br) $,
\end{itemize}
then it can be shown that
\begin{itemize}
\item[(a)]
\begin{equation} \label{eq:deltaUsc}
  \delta U^\s(\br) = \sum_{\s'=\al,\be} \int d\br' \;\Vsc(\br,\br')
  \;\tnosc(\br')
\end{equation}
which  immediately yields Eq.~(\ref{eq:DE2*}) from (\ref{eq:DE2}), and
\item[(b)]  
$\langle \delta  U_{ij}^2   \rangle  /   \Delta^2  $  and   
$\langle [\nKSosc^\s (\br) - \tnosc^\s (\br) ]^2 \rangle /  
\langle [\delta n^\s (\br)]^2 \rangle$ 
are both  negligible for a large dot for which the  chaotic motion in  the  potential
$\Utf^\s$ allows modeling of  the wave functions  in terms  of 
random  plane waves (being of
order $\ln g / g$ with $g$ the dimensionless
conductance of the dot).
\end{itemize}

\section{The Screened Interaction in the Local density approximation}
\label{sec:T&C}

The second order Strutinky corrections, Eqs.~(\ref{eq:DE2}) or
(\ref{eq:DE2*}), involve the bare and screened interactions
Eqs.~(\ref{eq:Vbare}) and (\ref{eq:Vsc}).  In this section, we shall
consider the particular form these interactions take for the
exchange correlation term we use to perform the actual SDFT
calculations, namely the local spin density approximation
\begin{equation} \label{eq:def_Exc}
\Exc[n^\al,n^\be] \simeq \int \!d\br \;n(\br) \;\exc\big(n(\br),\zeta(\br)\big) \; ,
\end{equation}
where $\zeta(\br) = (n^\al - n^\be)/n$ is the polarization of the
electron gas, and $\exc$ is the exchange plus correlation energy per
electron for the uniform electron gas with polarization $\zeta$.  We
furthermore use Tanatar and Ceperley's form of $\exc$ at $\zeta = 0$
and $1$,\cite{T&C} and the interpolation $\exc(n,\zeta) = \exc(n,0) +
f(\zeta) [\exc(n,1) - \exc(n,0)]$ for arbitrary polarization, with
$f(\zeta) = \left[ (1+\zeta)^{3/2} + (1-\zeta)^{3/2} - 2 \right] /
(2^{3/2} - 2)$.  Recently, Attaccalite et al.\cite{Attaccalite02}
parameterized a new LSDA exchange-correlation functional form based on
more accurate quantum Monte Carlo calculations (see e.g.\
Ref.~\onlinecite{Saarikoski03}) that include spin polarization
explicitly. We have, however, checked that for the quantities in which
we are interested here, this new functional introduces only minor
modifications; we shall therefore use Tanatar-Ceperley's form for our
discussion.

\begin{figure}
\includegraphics[height=2.6in]{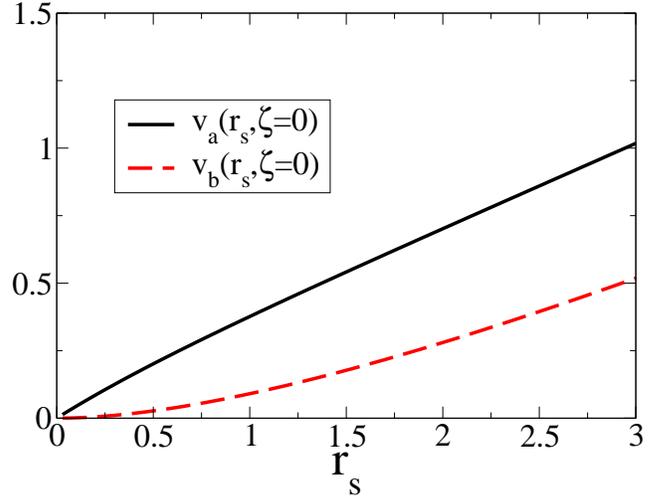}
\caption{(Color online) The functions  $v_a(n,0 )$  and $v_b(n,0)$, which  define the
  functional    second   derivative   of    $\Exc[n^\al,n^\be]$   [see
  Eq.~(\ref{eq:Vxc})],    as    a    function   of    the    parameter
  $r_s=1/\sqrt{\pi a_0^2 n}$.}
\label{fig:xc_coef}
\end{figure}

 From the  expression of  the functional $\Exc[n^\al,n^\be]$,  the bare
and     screened    interaction     potentials    Eq.~(\ref{eq:Vbare})
and(\ref{eq:Vsc}),  needed  for the  evaluation  of  the second-order
Strutinsky corrections, are easily  computed.  The bare interaction is
the  sum   of  two  terms  $\Vbare   =  \Vcoul  +   \Vxc$.  The
Coulomb interaction
\begin{equation}
  \Vcoul(\br,\br') = v_\mathrm{int}(|\br-\br'|)
\cdot \left( \begin{array}{cc} 1 & 1 \\ 1 & 1 \end{array} \right)
\end{equation}
is independent of both the density and spin. The matrix structure here
refers to the spin indices $\al$ and $\be$.  The exchange correlation
term is local and can be expressed as
\begin{eqnarray} \label{eq:Vxc}
  \lefteqn{\Vxc(\br,\br') =  - 2\pi a_0 e^2 \delta (\br - \br')} \qquad \\
& & \times
\left( \begin{array}{cc}
\va\big(n(\br),\zeta(\br) \big) & \vb\big((n(\br),\zeta(\br)\big)   \\
\vb\big(n(\br),\zeta(\br) \big) & \va\big(n(\br),-\zeta(\br) \big)
\end{array} \right) \; , \nonumber
\end{eqnarray}
where $\va$ and $\vb$ are obtained from the partial derivative of
$\exc(n,\zeta)$ [defined in Eq.~(\ref{eq:def_Exc})].  In all numerical
calculations we shall keep entirely the dependence of $\va$ and $\vb$
on the polarization $\zeta$.  However, this latter will usually be
relatively small, and $\va$ and $\vb$ contains already second
derivatives of the functional $\Exc$ with respect to the polarization.
We shall therefore proceed assuming $\va(n,\zeta) \simeq \va(n,0)$ and
$\vb(n,\zeta) \simeq \vb(n,0)$.  The dependence of these functions on
the parameter $r_s= (\pi a_0^2 n)^{-1/2}$, with $a_0 = \hbar^2/me^2$
the 2D Bohr radius, is shown in Fig~\ref{fig:xc_coef}.

Turning now to the screened interaction $\Vsc(\br,\br')$, it is useful
to switch  to the  variables $(\bR,\bl) =  ((\br'+\br)/2, (\br-\br'))$.
The Fourier transform of $\Vbare(\bR,\bl)$ with respect to $\bl$ reads
\begin{eqnarray} \label{eq:Vbare_II}
 \lefteqn{ \hat \Vbare(\bR,\bq) =
2 \pi e^2 a_0 v_c(\bq)
\cdot \left( \begin{array}{cc} 1 & 1 \\ 1 & 1 \end{array} \right) } \qquad\qquad \\
& & \mbox{ }
- 2\pi a_0 e^2
\left( \begin{array}{cc}
\va(\bR)  & \vb(\bR)   \\
\vb(\bR) & \va(\bR)
 \end{array} \right) \nonumber
\label{eq:hatVbare} 
\end{eqnarray}
where 
\begin{equation}
v_c(\bq) \equiv \frac{1}{a_0 |\bq|}(1-e^{-2 z_d q})
\end{equation}
(again, the pure Coulomb case is recovered by letting the distance to
the top gate $z_d$ go to infinity).  This can by further simplified by
diagonalizing the matrix in the spin indices: the eigenvectors are the
``charge channel'' $\vec c = (\vec\al + \vec\be)/\sqrt{2}$ and the
``spin channel'' $\vec s = (\vec\al - \vec\be)/\sqrt{2}$. The
eigenvalues are
\begin{equation}
  \lambda^{\rm bare}_c(\bR,\bq) =
  \frac{2}{N(0)}  \{2 v_c(q) - [\va(\bR) + \vb(\bR)]\}
\end{equation}
in the charge channel and
\begin{equation}
\lambda^{\rm bare}_s(\bR,\bq) = \frac{2}{N(0)} [\vb(\bR) -
  \va(\bR)]
\end{equation}
in the spin channel [note $2\pi a_0 e^2 = 2/N(0)$].

Because the screening length is  short on the scale
for  which   the  smooth  part   of  the  electronic   density  varies
appreciably, the operator $\delta^2 \Ttf  /\delta n^2 + \Vbare$ can be
inverted by treating the variable $\bR$ as a parameter, appearing thus
diagonal in the $\bq$ representation.
One obtains in this way
\begin{equation} \label{eq:hatVsc}
  \hat \Vsc(\bR,\bq) =
\left( \begin{array}{cc}
(\lambda_c+\lambda_s)/2 &  (\lambda_c-\lambda_s)/2   \\
(\lambda_c-\lambda_s)/2 &  (\lambda_c+\lambda_s)/2
 \end{array} \right) \; ,
\end{equation}
in terms of the eigenvalues
\begin{equation} \label{eq:lambda_c}
  \lambda_c(\bR,\bq)  = 
  \frac{2}{N(0)}  \frac{2 v_c(q) - [\va(\bR) + \vb(\bR)]}{
  1 + g(q) [2 v_c(q)  - (\va(\bR) + \vb(\bR))]} 
\end{equation}
for the charge channel and
\begin{equation} \label{eq:lambda_s}
\lambda_s(\bR,\bq)  =  \frac{2}{N(0)} \frac{[\vb(\bR) - \va(\bR)]}{1 + g(q)[\vb(\bR)
  - \va(\bR)]} 
\end{equation}
for the spin channel.
The $\bR$ dependence of $\lambda_c$ and $\lambda_s$ arises from
$\va(n(\bR))$ and $\vb(n(\bR))$, and therefore from the local value of
the density (that is, of the parameter $r_s$) at the location the
interaction is taking place.  Furthermore, the function
\begin{equation}
g(q) = \left( 1 + \eta \frac{q^2 }{8 \pi \nTF(\bR)} \right)^{-1}
\end{equation}
would  just  be   1  in  the  absence  of   the  $\hbar$  correction
$\Ttf^{(1)}$  to the  Thomas-Fermi  kinetic energy  term; it prevents
effective screening to take place for large momenta.

\section{The gated quartic oscillator model}
\label{sec:KSvsST}

To evaluate  the accuracy of  the Strutinsky approximation  scheme, we
consider  a two dimensional model system  for which  the electrons  are confined  by a
quartic potential
\begin{equation} \label{eq:qos}
  \Uext(x,y) = a \left[ \frac{x^4}{b} + b y^4 - 2\lambda x^2 y^2
    + \gamma(x^2y -y^2x)r \right] \; .
\end{equation}
The role of the various parameters of $\Uext(x,y)$ is the following:
$a$ controls the total electronic density (i.e.~the parameter $r_s$),
and therefore the relative strength of the Coulomb interaction;
$\lambda$ allows one to place the system in a regime where the
corresponding classical motion is essentially chaotic; finally, $b$
and $\gamma$ have been introduced to lower the symmetry of the system.
In the following sections, we use [a1] to denote the parameter value
$a=10^{-1} \times E_R/a_0^4$ (with $E_R = e^2/2a_0$), and [a4] for
$a=10^{-4} \times E_R/a_0^4$, which at $N=120$-$200$ correspond to
$r_s\simeq 0.3$ and $1.3$, respectively.  We use $\lambda=0.53$ and
$\gamma=0.2$ in our calculations unless specified otherwise.  In
addition to this 2-dimensional potential, we assume the existence of a
metallic gate some distance $z_d$ away from the 2D electron gas whose
purpose is to cut off the long-range part of the Coulomb interaction.
This gate is placed sufficiently far from the electrons to prevent the
formation of a density deficit in the center of the potential well
without modifying qualitatively the quantum fluctuation.  In practice,
we take $z_d$ about $0.75 \,a_0$ for [a1] , and $2.5 \,a_0$ for
[a4].\footnote{For the higher density case [a1], it was actually
necessary to place the gate quite close to the electron gas to see any
effect on the density, even on the classical scale.  The image charge
associated with the gate is at a distance $2 z_0$ which is, however,
still larger than the screening length within the electron gas.}

\subsection{Electronic densities}

For any set of parameters defining the potential Eq.~(\ref{eq:qos}),
and for any number of up and down electrons $(N^\al,N^\be)$, we can
compute the Kohn-Sham energies $\Eks[N^\al,N^\be]$ and densities
$\nKS^\s(\br)$ following the approach described in detail in
Ref.~\onlinecite{Jiang03b}.  For the Strutinsky approximation, the
only part which presents some degree of difficulty is actually the
Thomas-Fermi calculation, for which we have developed a new
conjugate-gradient method which turns out to be extremely
efficient.\cite{cgetf} Once $\nTF^{(\al,\be)}(\br)$ are known, the
effective potentials $\Utf^{(\al,\be)}(\br)$ and the corresponding
densities $\tn^{(\al,\be)}(\br)$ follow immediately.

\begin{figure}
\includegraphics[width=2.9in,clip]{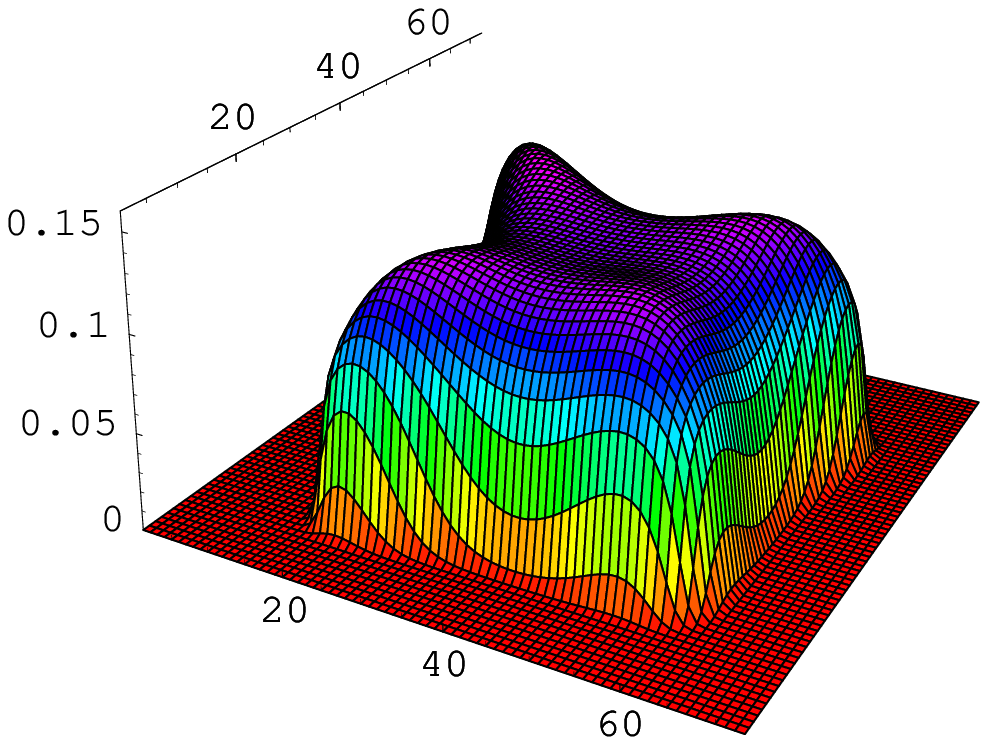}\vspace*{-0.3in}
\includegraphics[width=2.9in,clip]{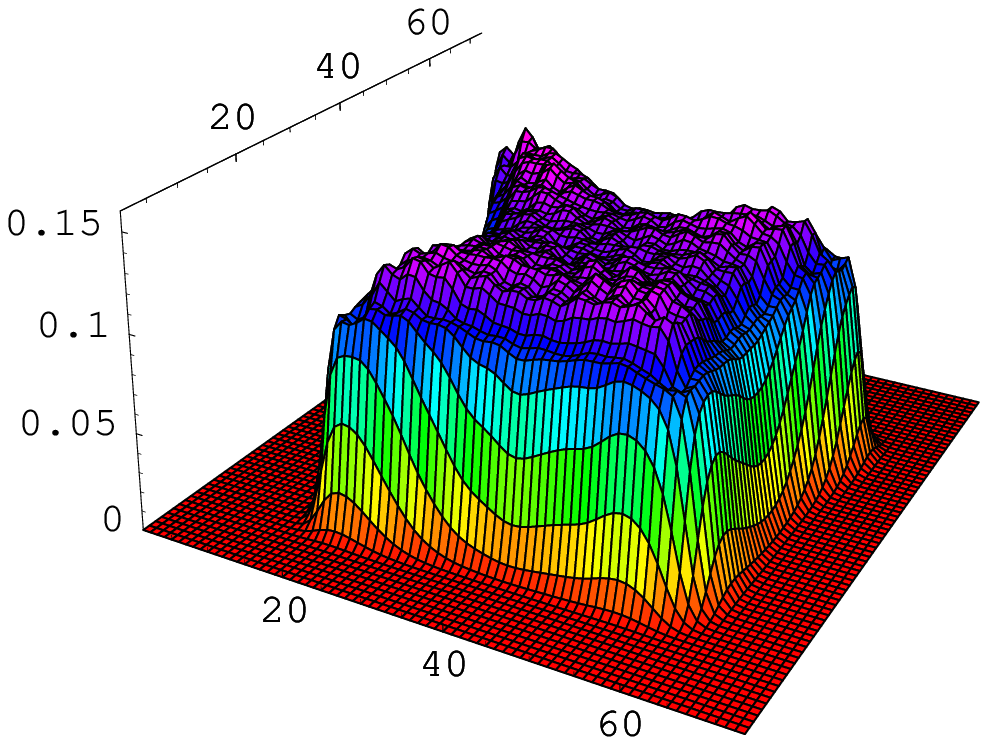}\vspace*{-0.3in}
\includegraphics[width=2.9in,clip]{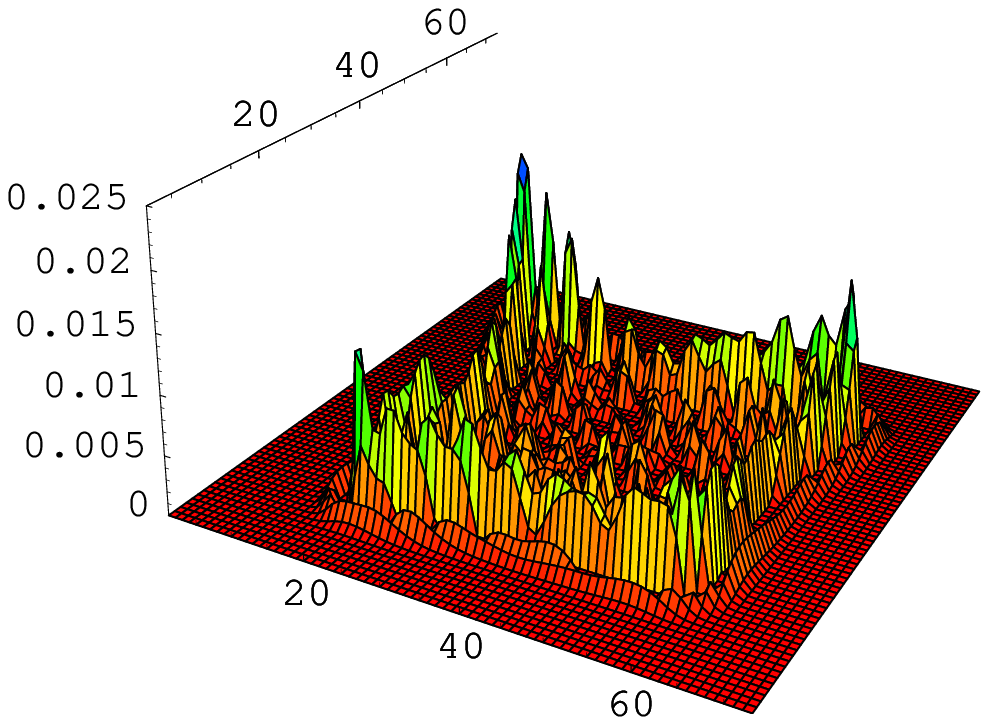}\vspace*{-0.3in}
\includegraphics[width=2.9in,clip]{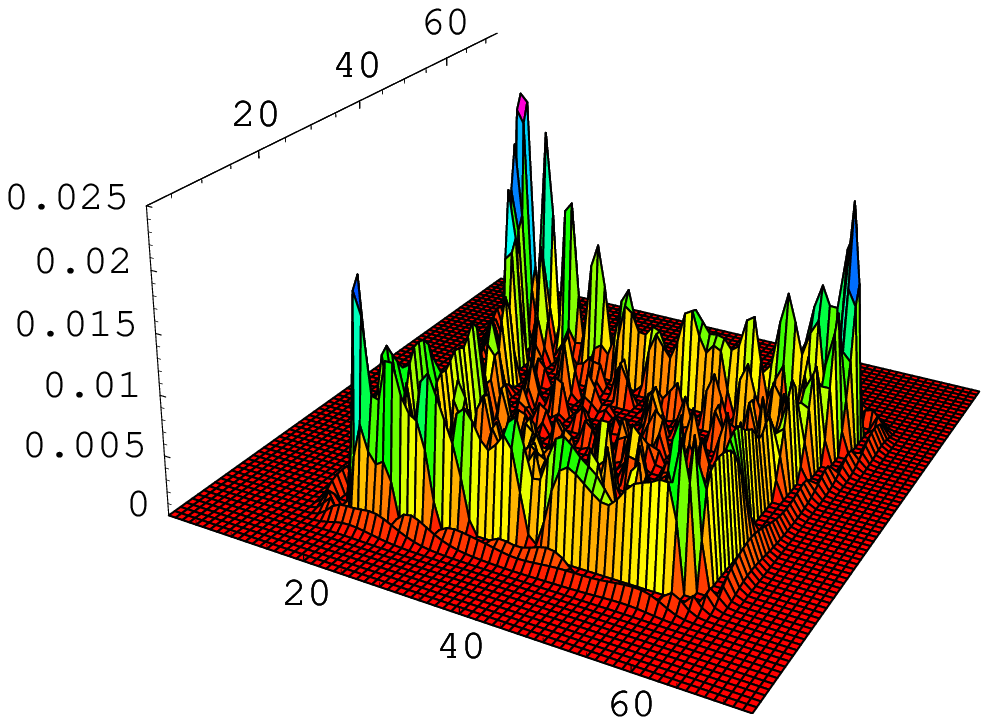}\vspace*{-0.1in}
\caption{(Color online) Particle  density for  the parameters  [a4] for  a  system of
$N=200$ electrons ($N_\al  = N_\be = 100$). From top to bottom: $\nTF(\br)$; 
$\nKS(\br)$;   $\delta  n(\br)$;  and
$\tnosc(\br)$. Note that $\nTF$ is a smooth approximation to $\nKS$
and that $\delta  n(\br)$ and $\tnosc(\br)$ are very similar.}
\label{fig:density_a4}
\end{figure}

\begin{figure}
\includegraphics[width=2.8in,clip]{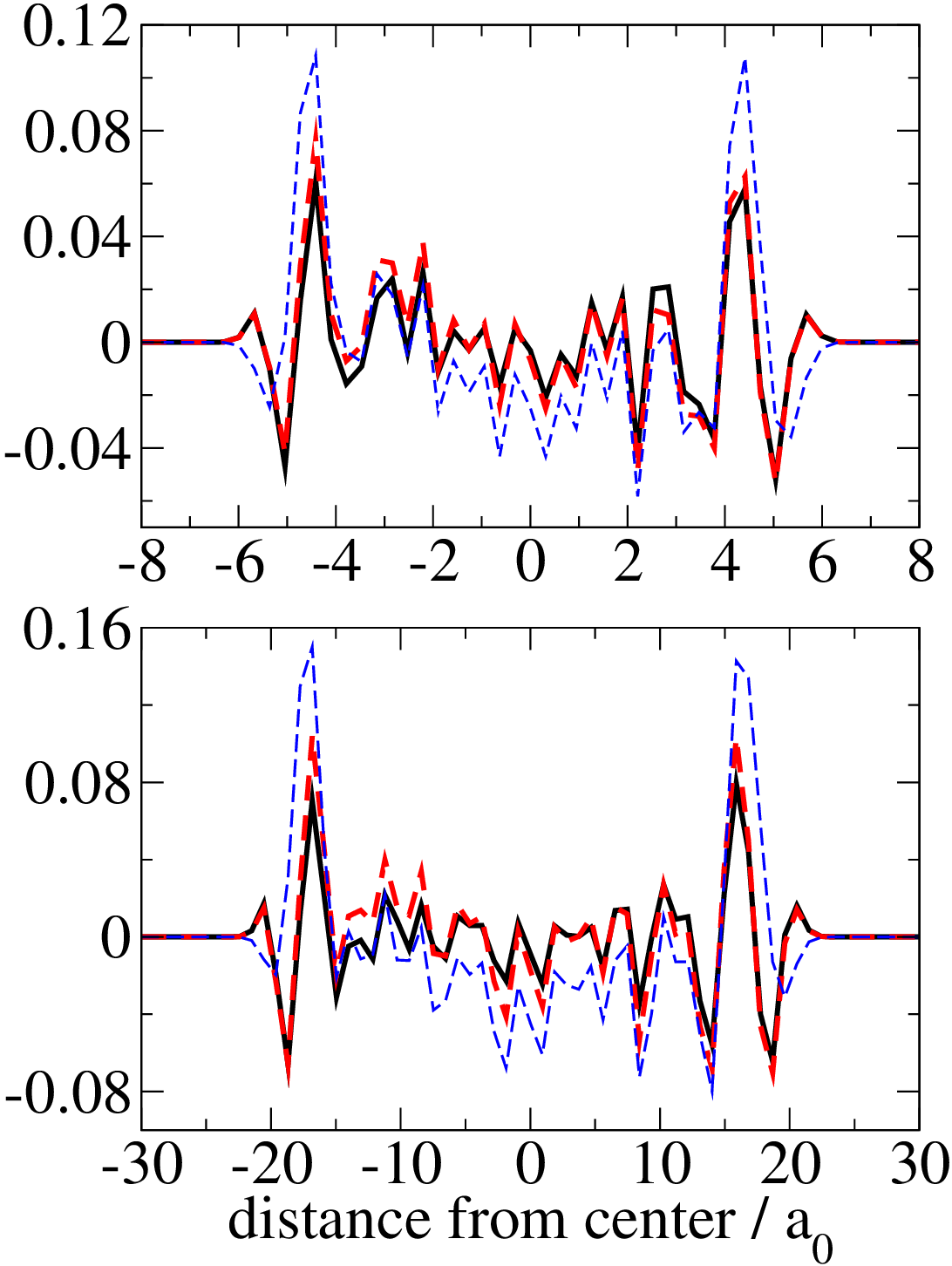}
\caption{(Color online) Diagonal cut of particle densities $\delta
n(\br)$ (solid) and $\tnosc(\br)$ (dashed), in units of the average
Thomas-Fermi density $\nTF$ inside the dot.  Upper panel: parameters
[a1] ($r_s \simeq 0.3$); Lower panel parameters [a4] ($r_s \simeq
1.3$).  The dot contains $N=200$ electrons ($N_\al = N_\be =
100$). The thin dashed lines correspond to $\tnosc(\br)$ from a less
accurate choice of the Thomas-Fermi kinetic energy term, namely
$\eta=1.0$ instead of $0.25$.}
\label{fig:diagonal_cut}
\end{figure}

Figure~\ref{fig:density_a4} shows the densities $\nTF(\br)$,
$\nKS(\br)$, $\delta n(\br)$, and $\tnosc(\br)$ for the ground state
with $N=200$ electrons ($N_\al = N_\be = 100$) of the gated quartic
oscillator system with parameter [a4], corresponding to an interaction
parameter of $r_s = 1.3$.  Noting in particular the difference of
scale between the upper and lower panels, one can observe that the
Thomas-Fermi density already is a very good approximation to the exact
one, and that $\delta n(\br)$ is a small oscillating correction, of
the same magnitude as the oscillating part of $\tn(\br)$.  Apparent
also on this figure is the fact that the largest errors are located at
the boundary of the dot where corrections to the Weyl density of
higher order in $\hbar$ are the largest.  To make this more visible,
we plot in Fig.~\ref{fig:diagonal_cut} the densities $\delta n(\br)$
and $\tnosc(\br)$ along a cut on a diagonal of the confining potential
for two sets of parameters.  This makes it clear that $\tnosc(\br)$ is
an oscillating function only in the interior of the dot, but has a
proportionally large secular component at the boundary.  As a
consequence, choosing correctly the term $\Ttf^{(1)}$ of the
Thomas-Fermi kinetic energy term is actually important to obtain good
accuracy.  We have therefore determined the parameter $\eta=0.25$ of
this functional by imposing that $\Eosc[N]$ oscillates around zero,
rather than having a significant mean value.  As an illustration, we
also plot in Fig.~\ref{fig:diagonal_cut} the same quantities but for a
calculation where a value $\eta=1$ has been used for the Thomas-Fermi
kinetic energy correcting term.  We see that this increases
significantly the error in the Thomas-Fermi density at the boundary,
reducing the effectiveness of the Strutinsky approximation.

\subsection{Ground state energies}

With the expressions Eqs.~(\ref{eq:Vbare_II}) and (\ref{eq:hatVsc}) of
the bare and screened interactions, and using the known eigenvalues
and eigenfunctions for the Schr\"odinger equations Eqs.~(\ref{eq:KS})
and (\ref{eq:schroe}), the Strutinsky approximation for the total
energy, including the order one Eq.~(\ref{eq:DE1}) and order two
Eqs.~(\ref{eq:DE2}) or (\ref{eq:DE2*}) corrections, can be computed
for any $(N_\al,N_\be)$.  As for the full spin density calculations,
the ground state energy for a given total number of electron $N$ is
then obtain as the minimum over $(N_\al,N_\be)$ with $N_\al + N_\be =
N$ of these energies.

\begin{figure}
\includegraphics[width=2.8in]{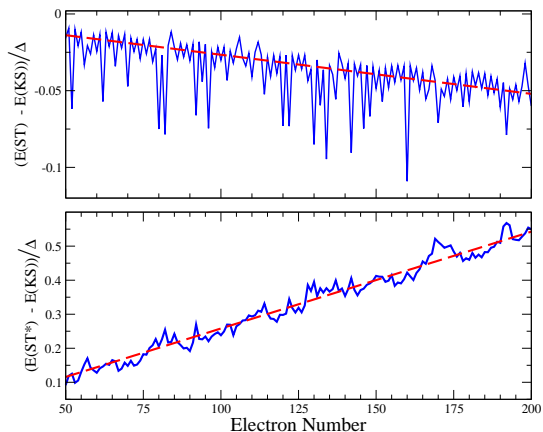}
\caption{(Color online) Difference between Strutinsky and Kohn-Sham ground state
energies, in units of the mean level spacing, as a function of the
number of electrons, for the high density dot [a1] ($r_s \simeq 0.3$).
Upper panel: $\Est[N]$ obtained from Eq.~(\ref{eq:DE2}); Lower panel:
$\Eststar [N]$ obtained from Eq.~(\ref{eq:DE2*}). In both cases, the
difference shows small fluctuations (a few percent) about a linear
secular trend (dashed line is fit).}
 \label{fig:EST-EKS:1}
\end{figure}

Let  us  now consider  these  ground state  energies  for  a choice  of
parameters [a1], such that  the coefficient $r_s=0.3$ is still smaller
than one and thus the  Coulomb interaction is not very large compared
to  the  kinetic energy  of  the particles.   Fig.~\ref{fig:EST-EKS:1}
displays the corresponding difference between Strutinsky and Kohn-Sham
ground-state energies,  in mean level spacing units,  for a number of
electrons ranging from  50 to 200.  For the  upper panel, $\Est[N]$ is
obtained  from  the   expression  Eq.~(\ref{eq:DE2})  using  the  bare
interaction  and   requiring  the  knowledge  of   the  exact  $\delta
n^{(\al,\be)}$.   For the  lower panel,  $\Eststar[N]$ is  obtained from
Eq.~(\ref{eq:DE2}), which involves  the approximate screened interaction, but only
the knowledge of $\tnosc^{(\al,\be)}(\br)$.

The first observation that can be made on that figure is that the
second form of the Strutinsky approximation appears substantially less
accurate than the first one.  As mentioned earlier, this is, however,
probably due to the fact that, because our code was devised to handle
only two body interactions that were diagonal either in position or in
momentum representation, we had to use in that calculation an
approximation where, for the screened interaction, the local value of
the parameter $r_s(\br) = \sqrt{1/\pi \nTF(\br)}$ has been replaced by
its mean value .

Indeed, a second feature immediately visible on
Fig.~\ref{fig:EST-EKS:1} is the presence of a net trend in the energy
differences between Kohn-Sham and Strutinsky calculations.  This
secular term is related to the non-oscillating component of
$\tnosc(\br)$ and $\delta n(\br)$ visible on the lower panels of
Figs. \ref{fig:density_a4} and \ref{fig:diagonal_cut} at the boundary
of the quantum dots.  This can be checked by using a less accurate
Thomas-Fermi approximation (e.g. with $\eta=1$ for the correcting term
of the kinetic energy $\Ttf^{(1)}$), and noticing that this secular
term in the deviation increases noticeably while the fluctuations
remain less affected.  Since the total density of electrons at the
boundary of the dot is significantly lower than its average value, it
is relatively natural that in our approach, this secular deviation is
made significantly worse in the second form of our approximation.

\begin{figure}
\includegraphics[width=2.8in]{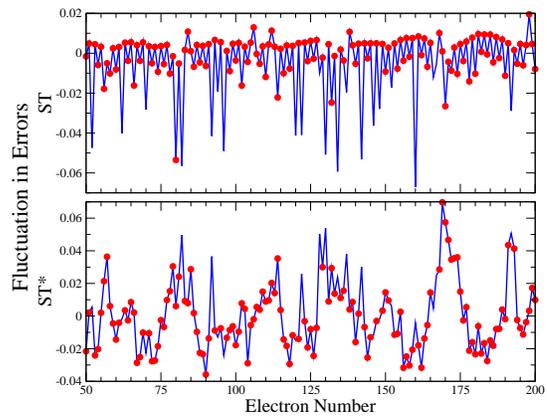}
\caption{(Color online) Difference between Strutinsky and Kohn-Sham ground state
energies as in Fig.~\ref{fig:EST-EKS:1} but now with the secular trend
removed.  Solid: all ground states; Dots: ground states without
significant spin contamination. Note the excellent agreement in the
case of the first Strutinsky form $\Est$ (upper panel) -- of order 1
percent -- when spin contamination is not present.  In the case of the
second form $\Eststar$, the agreement is still very good (lower
panel).  }
 \label{fig:EST-EKS:2}
\end{figure}

The secular deviation is not completely negligible in terms of the
mean level spacing.  The relevant scale for the smooth part of the
ground energies is, however, the charging energy, compared to which
these secular corrections are extremely small.  To focus on the
fluctuating part, we therefore remove the secular term (i.e. the
straight lines in Fig.~\ref{fig:EST-EKS:1}), obtaining in this way
Fig.~\ref{fig:EST-EKS:2}.

Let us consider first the upper panel, where the first form of the
Strutinsky approximation has been used.  We observe that the
fluctuating part of the error is usually extremely small, typically of
order a percent of a mean level spacing, and that this is mainly an
oscillation between odd and even number of particles in the system.
Nevertheless, in a few circumstances significantly larger deviations
are observed, with a magnitude typically five percent of a mean level
spacing and a sign which is always negative.

\begin{figure}
\includegraphics[width=2.8in]{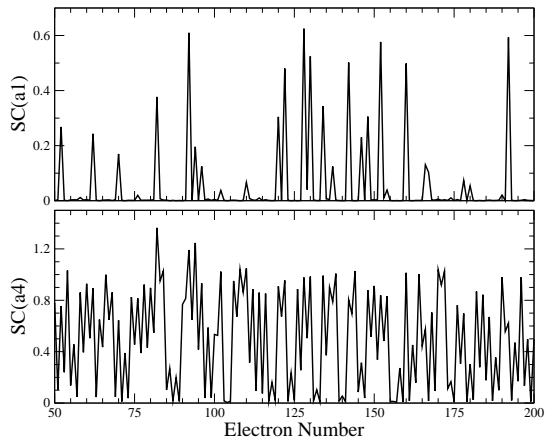}
\caption{Spin contamination $\delta S[N]$ as a function of the number
  of particles. The spin contamination is infrequent at high density
  ([a1], upper panel), but becomes frequent and substantial for $r_s
  \simeq 1.3$ ([a4], lower panel).}
 \label{fig:SC}
\end{figure}

To understand the  origin of these larger deviations,  it is useful to
correlate them  with the occurrence  of spin contamination, that  is to
situations where the  SDFT calculations  break the  spin rotation
symmetry.   Since the  actual spin  is a  somewhat ill-defined
quantity  in a  spin density  calculation,  we need  however first  to
specify what we understand by this.  Indeed, in spin density
functional theory the difference $N_\al - N_\be$ can be interpreted as
twice the component $S_z$ of the quantum  dot total spin. Another
quantity that can be easily computed is the mean value $S(S+1)$ of the
operator  $\vec  S^2$  for  the  Slater  determinant  formed  by  the
Kohn-Sham  orbitals $\psi^{\sigma}_i$,  $ i  = 1,\cdots,N_{\sigma}$,
which can  be expressed as $S(S+1)  = S_z(S_z+1) + \delta  S$ with the
``spin contamination'' $\delta S$ given by\cite{spin_contamination}
\begin{equation}
   \delta S =  N_\be - \sum_{i,j \le N_\be}
  | \langle \psi^\al_i \vert \psi^\be_j \rangle |^2 \; .
\end{equation}
 From this expression, one sees that if all occupied $\be$-orbitals are
identical to the corresponding $\al$-orbitals, $\delta S = 0$, and it
is presumably reasonable to interpret $(N_\al - N_\be)/2$ as the
system's total spin.  However, as soon as different-spin orbitals
start to differ, $\delta S$ can take any positive value smaller than
$N_\be$, signaling that, at least, there is some ambiguity in the
assessment of the total spin of the system.  In Fig.~\ref{fig:SC} we
have plotted the ground state spin contamination $\delta S [N]$ as a
function of the particle number. For [a1], the spin contamination is
usually negligible, except in some few cases where $\delta
S$'s of order one half or so are encountered.

Coming back to the Strutinsky approximation, we can use the
information from Fig.~\ref{fig:SC} to exclude in
Fig.~\ref{fig:EST-EKS:2} the ground states with significant spin
contamination.  The remaining points correspond to the dotted symbols
in this figure. In the upper panel, we see that there is almost a one
to one correspondence between larger errors and spin contamination.

Turning to the lower panel in Fig.~\ref{fig:EST-EKS:2}, we see that
the further approximations in treating the screening used in
evaluating Eq.~(\ref{eq:DE2}) do degrade the accuracy of the ground
state energy somewhat. Still, ST* gives the fluctuating part of the
energy to within a few percent. For the spin contamination, no
particular correlation is seen, presumably again because the overall
agreement is slightly spoiled by the approximation we made for the
screened Coulomb interaction.

\begin{figure}
\includegraphics[width=2.8in]{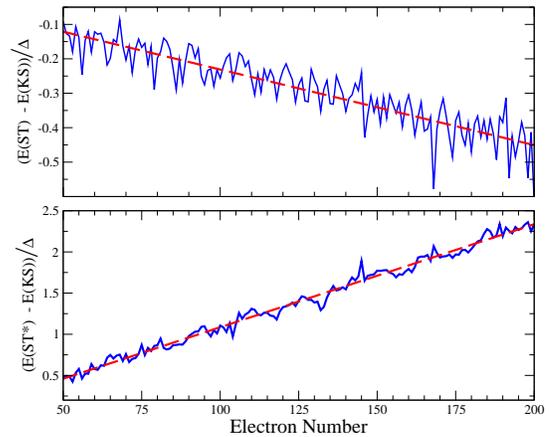}
\caption{(Color online) Difference between Strutinsky and Kohn-Sham ground state
energies, in units of the mean level spacing, as a function of the
number of electrons, for the $r_s \simeq 1.3$ dot [a4].  Upper panel:
$\Est[N]$; Lower panel: $\Eststar [N]$.  The linear secular trend
(dashed line is a fit) is now significantly larger than for [a1]
(compare with Fig. \ref{fig:EST-EKS:1}), but the fluctuation about
this trend remains small (of order 5 percent).}
 \label{fig:EST-EKS-a4}
\end{figure}

In lower density (larger $r_s$) more realistic dots modeled by the
parameter set [a4], spin contamination in KS ground states is much
more pronounced, as shown in the lower panel of Fig.~\ref{fig:SC} --
it is, in fact, always significant. In conjunction,
Fig.~\ref{fig:EST-EKS-a4} shows that the accuracy of both ST [i.e.\
using Eq.~(\ref{eq:DE1})] and ST* [i.e.\ using Eq.~(\ref{eq:DE2})]
becomes worse.  As at higher density, the main error is a secular
trend: in the case of ST* it attains a value of several mean level
spacings, due presumably to the approximations made in treating the
screening. After removing the secular deviation, however, the
fluctuation in the errors of ST and ST* ground state energy is still
quite small: the rms is $0.05 \Delta$ in ST and $0.06\Delta$ in
ST*. Thus for characterizing the fluctuating part of the ground state
energy, ST* is nearly as good as ST, a property we expect to remain
valid at larger $r_s$.

\subsection{Coulomb Blockade peak spacings and spin distribution}
\label{sec:RPW}

In the previous subsection, we have considered the accuracy of the
Strutinsky approximation for individual ground state energies.  We
found that as long as no significant amount of spin contamination is
present in the SDFT calculation, the Strutinsky result provides an
excellent approximation when Eq.~(\ref{eq:DE2}) is used, and a good,
though slightly degraded one, with Eq.~(\ref{eq:DE2*}).  In this
latter case, it is probable that the additional errors come mainly
from the neglect of the local density dependence in the screened
interaction rather than to the Strutinsky approximation itself.  We
shall come back to the issue of spin contamination in the next
section, and sketch an extension of the theory that would make it
suitable to deal with the spin contaminated case.

Before doing so, however, we shall address another question, namely
how well, even in the case where a one-to-one comparison of ground
state energies can imply an error of a fraction of mean level spacing,
are the statistical properties of the quantum dots described within
the Strutinsky approximation.  For instance, we have in mind the
distribution of ground state spin $S_z[N]$ or of ground state energy
second difference $s[N] = \Eks[N+1] + \Eks[N-1] - 2\Eks[N+1]$.  This
latter quantity is accessible experimentally by measuring the spacing
between conductance peaks in the Coulomb Blockade regime, and will be
referred to below as the ``peak spacing''.

\begin{figure}
\includegraphics[width=3.2in]{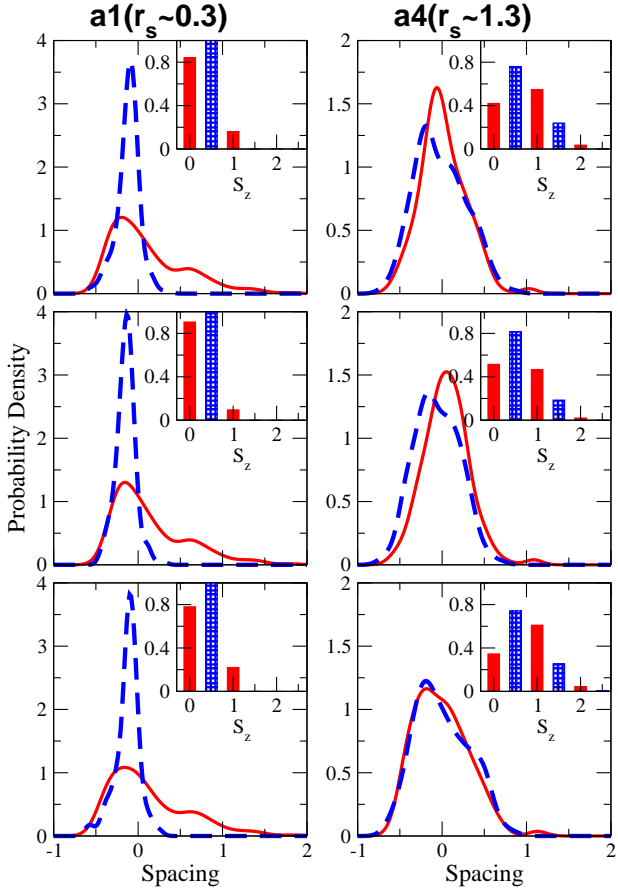}
\caption{(Color online) Spin and peak spacing distributions for the
  cases [a1] (left column) and [a4] (right column). Solid: even $N$;
  Dashed: odd $N$. The statistics are obtained for $N\!=\!120$-$200$ with
  $(\lambda,\gamma)\!=\!(0.53,0.2), (0.565,0.2), (0.6, 0.1), (0.635,
  0.15)$ and $(0.67, 0.1)$.  From top to bottom: Kohn-Sham , first,
  and second form of the Strutinsky approximation.  Agreement between
  the three methods is excellent, of course, for [a1]. But even for
  [a4] where individual energies may be in error, the agreement of the
  distributions is very good.  }
\label{fig:psp}\end{figure}

In Fig.~\ref{fig:psp}, both peak spacing and spin distributions are
plotted for two interaction strength regime ([a1] and [a4]) using
either Kohn-Sham results or one or the other form of the Strutinsky
approximation.  In the small $r_s$ case, the agreement is naturally
excellent, but we see that even for the higher $r_s$ case, both forms
of the Strutinsky approximation give a fairly good approximation --
certainly they provide a qualitatively correct description.

\section{Fermi liquid picture}
\label{sec:FL}

It is not possible to develop a real Landau Fermi liquid theory for
quantum dots because the mesoscopic fluctuations prevent any Taylor
expansion of the free energy in terms of occupation number.  (Landau
theory basically assumes that the excitation energies are the smallest
energy scales of the problem, which is clearly not true for mesoscopic
systems because of variation on the scale of the mean level spacing.)
However, discussing what we may call a Landau Fermi liquid
``picture'', in the sense that the low energy physics is described by
a renormalized weak interaction, is still something meaningful.

In that sense, what Eqs.~(\ref{eq:strut_corr})-(\ref{eq:Vsc})
express is that SDFT, in the limit where the Strutinsky approximation
scheme is accurate, is equivalent to a Landau Fermi liquid
picture, where quasiparticles with spin $\s=\al,\be$ evolve in the
effective potential $\Utf^\s$ and interact through a residual weak
interaction $\Vsc^{\s,\s'}(\br,\br')$ that can be taken into account
as a perturbation.  The only unusual feature is the absence of an
exchange-like contribution to the total energy.  Indeed the main role
of the exchange correlation functional $\Exc[n_\al,n_\be]$ is to make
the interaction between same spin particles different from the one
between opposite spins.

Since moreover we have chosen the confining potential in such a way
that the classical motion within our model quantum dot is chaotic, we
know that we can use a statistical description of the eigenlevels and
eigenstates of $\Htf = \bp^2/2m + \Utf(\br)$ in terms of Random Matrix
Theory (RMT) and Random Plane Wave (RPW) modeling.  We are therefore
in the position to follow the line of reasoning in
Ref.~\onlinecite{prb_ullmo_01} to analyse the SDFT calculation.  We
shall do this in this section first to get some understanding of the
peak spacing and spin distributions obtained, and in a later stage to address
the mechanism of spin contamination.

\subsection{Universal Hamiltonian form}

To model the statistical properties obtained from
the SDFT calculations, let us assume that the Thomas-Fermi density
across the dot has variation small enough that we can take the
parameter $r_s$ as a constant.  We furthermore impose $\nTF^\al(\br) =
\nTF^\be(\br) = \nTF(\br)/2$ ($\int \nTF^{\sigma}(\br) dr$ might then
be half integer for odd $N$, which is not a problem since at the
Thomas-Fermi level the quantization of particle number is not playing
any role) and write the second order Strutinsky correction as
\begin{equation} \label{eq:DE2_RPW}
  \Delta E^{(2)} = \frac{1}{2} \sum_{\s,\s' \atop i,j}
  f_{i \s} f_{j \s'} M^{\s,\s'}_{i,j} \,  - \,\overline{\Delta E^{(2)}} \; ,
\end{equation}
with $f_{i\s} = 0,\,1$ the occupation number of orbital $i$ with spin
$\s$, 
\begin{equation}
\overline{\Delta E^{(2)}} = \frac{1}{2} \sum_{\s,\s'} \int \!d\br
d\br'\;\nTF^\s(\br) \;\Vsc^{\s,\s'}(\br,\br') \;\nTF^{\s'}(\br') \;, 
\end{equation}
and
\begin{equation}
  M^{\s,\s'}_{i,j} = \int \!d\br d\br' \,|\phi_i(\br)|^2 \;\Vsc^{\s,\s'}(\br,\br')
  \; |\phi_j(\br')|^2 \; .
\end{equation}

For a  chaotic system, it  can be shown  that the fluctuations  of the
$M^{\s,\s'}_{i,j}$ are small (variance of order $\sim
\ln g/g^2$) and that their mean values are given by
\begin{equation} \label{eq:meanMij}
  \langle M^{\s,\s'}_{i,j}  \rangle = \Big[ \hat  \Vsc^{\s,\s'}(q=0) + \mu_{\rm T}
  \delta_{ij} \langle \Vsc^{\s,\s'}\rangle_{\rm fc} \Big] / A \; ,
\end{equation}
where $A$ is  the area of the  dot, $\mu_{\rm T}$ is 2 here
since time reversal symmetry is preserved  (but would be 1 if it were
broken), and
\begin{equation} \label{eq:mean_fc}
\langle \Vsc^{\s,\s'}\rangle_{\rm fc} = \frac{1}{2\pi} \int_0^{2\pi} \!\!d\theta
\;\hat \Vsc^{\s,\s'} \big(\sqrt{2(1+\cos\theta)}\kf\big)
\end{equation}
is the average of the screened interaction over the Fermi circle (note
$a_0 \kf = \sqrt{2} / r_s$).

Note however that the screened interaction
Eq.~(\ref{eq:Vsc}) is derived under the assumption that the
oscillating part of the density integrates to zero, so that the total
displaced charge providing the screening also does.  Between the
reference $S=0$ configuration (the TF case) and the higher $S$ ones,
the total number of electrons is, of course, conserved, but not the
number for each spin.  It should be kept in mind, therefore, that the
$q=0$ component of the density cannot be screened; this can be
included simply by setting $\lambda_s (q=0) \equiv \lambda_s^{\rm
bare}(q=0)$.

\begin{figure}
\includegraphics[height=2.6in]{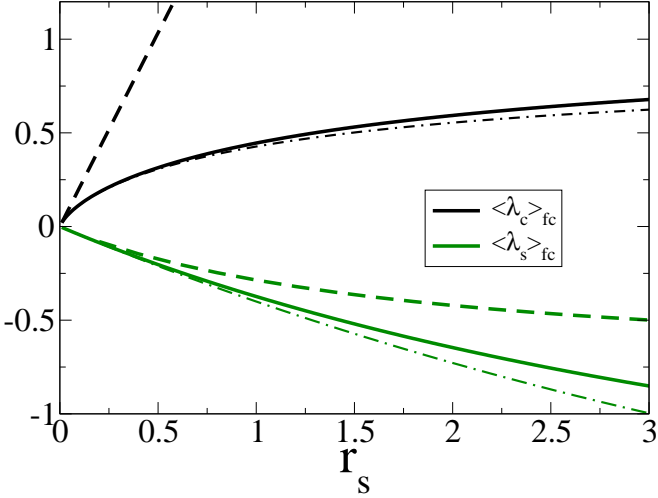}
\caption{(Color online) Average over the Fermi circle of the
eigenvalues of the screened and bare SDFT interactions [in units of
$2/N(0)$] as a function of $r_s=(\pi a_0^2 n)^{-1/2}$.  Dark: charge
channel. Lighter color (green online): spin channel.  Dashed: bare
interaction $\langle \lambda^{\rm bare}_{c,s} \rangle$ (with a cutoff
of the momentum at $q \simeq 1/L$ in the charge channel). Solid:
screened interaction $\langle \lambda_{c,s} \rangle$ with
$\eta=0.25$. Thin dot-dashed: same but with $\eta =0$.  Since the $q$
dependence of $\lambda_{s}$ is entirely due to the $\Ttf^{(1)}$
correction to the Thomas-Fermi kinetic energy functional, it therefore
disappears in this latter case. The interaction in the charge channel
is, of course, dramatically decreased by screening; in contrast,
screening increases the magnitude of the interaction in the spin
channel.  }
 \label{fig:lambda_cs}
\end{figure}

The screened interaction matrix is characterized by its eigenvalues
Eqs.~(\ref{eq:lambda_c}) and (\ref{eq:lambda_s}).  In
Fig.~\ref{fig:lambda_cs}, we thus plot the $r_s$ dependence of these
quantities averaged over the Fermi circle, and compare them to their bare
counterparts. 

A few remarks are in order concerning this figure.  First, we see
that, because of the divergence of the Coulomb interaction at small
$q$, screening has a drastic effect for the charge
channel.\footnote{Actually $\langle \lambda^{\rm bare}_c \rangle_{\rm
fc} $ would be divergent if we did not use a cutoff on $q$ at the
inverse of the system size.} Screening is less dramatic in the spin
channel, but can still be an order one effect as $r_s$ increases.

Furthermore, while the screening decreases the absolute strength of
the interaction in the charge channel, it actually {\em increases} it
in the spin channel.  Indeed, since the interaction in the spin
channel coming from SDFT is attractive, the charges in the bulk of the
Fermi sea will, as long as this does not involve too much kinetic
energy, move so as to increase the spin polarization.

Finally, we note that for the value of the parameter $\eta$ that we
use, the effect of the first $\hbar$ correction $\Ttf^{(1)}$ on the
screened interaction is very small in the charge channel, and only slightly
larger in the spin channel.

\begin{figure}
\includegraphics[width=2.8in]{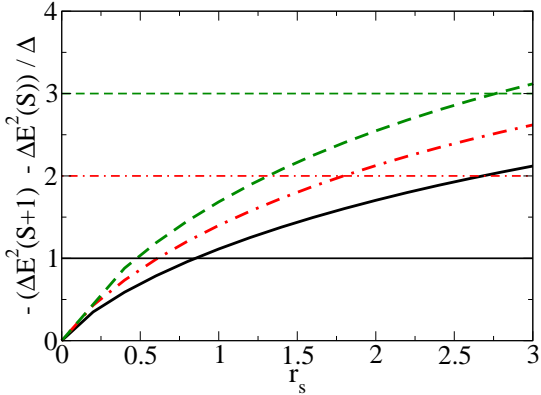}
\caption{(Color online) Mean value of the gain in interaction energy
  $- \left( \Delta E^{(2)} (S_z+1) - \Delta E^{(2)} (S_z) \right)$
  (thick lines) and of the mean one-particle energy cost (thin
  horizontal lines) associated with flipping the spin of one particle
  in the quantum dot. Solid: $S_z=0$; Dot-Dashed $S_z=1/2$; Dashed:
  $S_z=1$}
\label{fig:DE2(S)}
\end{figure}

If   we  neglect   the   fluctuations   of   the   $M^{\s,\s'}_{i,j}$,
Eqs.~(\ref{eq:hatVsc}), (\ref{eq:DE2_RPW}),  and (\ref{eq:meanMij}) imply
that $\Delta E^{(2)}$ is just a function of the number of particles $N
=  N_\al + N_\be$  in the  dot and the  $z$-component $S_z=(N_\al
- N_\be)/2$ of the groundstate   the spin,
\begin{eqnarray} \label{eq:DE2_UH}
  \lefteqn{ \Delta E^{(2)}(N,S_z) = \frac{1}{2} \mu_{\rm T} N \langle
\lambda_c \rangle_{\rm fc} + \lambda^{\rm bare}_s S_z^2 } \qquad
\qquad \qquad \qquad\\ & & \mbox{} - \frac{\mu_{\rm T}}{2}
\big(\langle \lambda_c \rangle_{\rm fc} - \langle \lambda_s
\rangle_{\rm fc} \big)S_z \; .  \nonumber
\end{eqnarray}
The main value of this expression is how it compares to the universal
Hamiltonian form,\cite{KurtlandPRB00,ABG02} and we shall come back to
this point in the discussion section.  Already we can see, however,
that it contains almost all the information necessary to understand
qualitatively the ground-state spin distributions. Indeed, looking at
Fig.~\ref{fig:DE2(S)}, which shows the difference $- \left( \Delta
E^{(2)}(N,S_z+1) - \Delta E^{(2)}(N,S_z) \right)$ as a function of
$r_s$ for several values of $S_z$, we see that for $r_s \simeq 0.85$,
the interaction energy gain and one particle energy cost of forming a
triplet are equal on average, and therefore triplets should become as
probable as singlets.  In the same way, spin 3/2 becomes as probable
as 1/2 at $r_s \simeq 1.8$, and spin 2 as probable as 1 at $r_s \simeq
2.8$.

\begin{figure}[t]
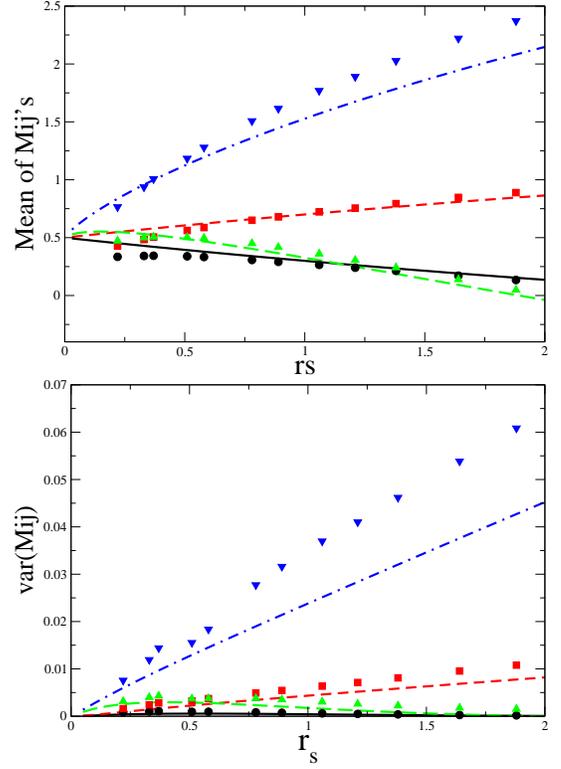

\includegraphics[width=2.8in,clip]{Mij_mean.eps}
\includegraphics[width=2.8in,clip]{Mij_var.eps}
\caption{(Color online) Comparison between the analytical RPW
predictions [Eqs.~(\ref{eq:meanMij}) and (\ref{eq:varMij})] and
numerical calculations of the mean and variance of $M_{ij}$'s. The
wave functions used in the numerical calculations are eigenfunctions
of the effective Thomas-Fermi potential with $N=200$ in the quartic
oscillator system. Lines (points) correspond to analytical (numerical)
results, with the solid (circle) for $M_{i,j}^{\sigma,\sigma}$,
short-dashed (square) for $M_{i,j}^{\alpha,\beta}$, long-dashed
(up-triangle) for $M_{i,i}^{\sigma,\sigma}$ and dot-dashed
(down-triangle) for $M_{i,i}^{\alpha,\beta}$. }
\label{fig:Mijs}
\end{figure}

To have a more precise idea of the whether the random-plane-wave model
captures the main physics, we can follow the approach of
Ref.~\onlinecite{prb_ullmo_01} and make a simulation of the peak
spacing and spin distributions. We use GOE distributed energy levels
for the first-order correction Eq.~(\ref{eq:DE1}), and take the
second-order correction in the form Eq.~(\ref{eq:DE2_RPW}) with the
$M^{\s,\s'}_{i,j}$ independent variables with mean
Eq.~(\ref{eq:meanMij}) and a variance which can be computed using the
method of Appendix~A of Ref.~\onlinecite{prb_ullmo_01}:
\begin{eqnarray} \label{eq:varMij}
  \lefteqn{ \frac{ {\rm var}\big[M^{\s\s'}_{i,j} \big] }{\Delta^2}
 = \frac{32}{\pi (\kf L)^2}
  \int_{\frac{2\pi}{\kf L}}^{2- \frac{2\pi}{\kf L}} \frac{dx}{x}
  \frac{1}{4-x^2}   \hat v^{\s\s'}(x)           
  } \\
  & & \times \left[ \hat v^{\s\s'}(x)
  + \delta_{ij} \Big(\hat v^{\s\s'}(x) + \hat v^{\s\s'}(0) +
\hat v^{\s\s'}\big(\sqrt{4-x^2}\big) \Big) \right]  \nonumber
\end{eqnarray}
where $\hat v^{\s\s'}(q/\kf) \equiv N(0) \hat \Vsc^{\s \s'}(q)/2$.
That the random-plane-wave model correctly describe the wave function
statistics is illustrated in Fig.~\ref{fig:Mijs} where the analytic
expressions Eqs.~(\ref{eq:meanMij}) and (\ref{eq:varMij}) are
compared, as a function of $r_s$, to the result from the actual
eigenfunctions derived from Eq.~(\ref{eq:schroe}). The analytic
expression for the mean is expected to be quite reliable and hence the
good agreement. The variance (\ref{eq:varMij}) is
less accurate -- because of the cutoff used, for instance -- and so we
consider the agreement in Fig.~\ref{fig:Mijs} quite good.

\begin{figure}
\includegraphics[width=3.2in,clip]{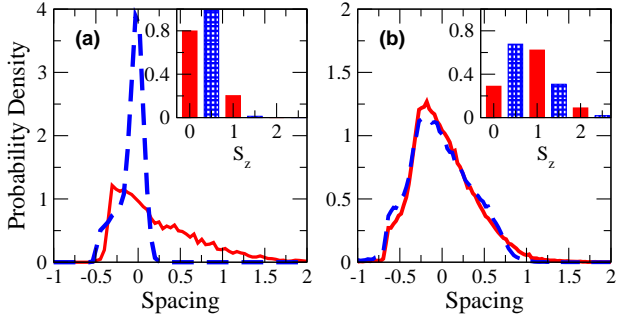}
\caption{(Color online) Peak spacing distributions for the RMT/RPW model
  corresponding to $r_s = 0.3$ (left) and $1.3$ (right). Solid~: even $N$,
  dashed~: odd $N$. Inset~:
  corresponding spin distribution. }
\label{fig:level4}
\end{figure}

Fig.~\ref{fig:level4} displays the peak spacing and spin distribution
for $r_s = 0.3$ (corresponding to [a1]) and $r_s = 1.3$ ([a4]) coming
from a simulation in which the fluctuations of level spacing and of
the $M^{\s,\s'}_{i,j}$ are included.  We see that the qualitative
behavior observed in Fig~.\ref{fig:psp} -- and in particular the lower
panel -- is very well reproduced. Thus the RMT/RPW approach
using the LSDA interaction is successful in comparison with the full
SDFT calculation.

\subsection{Spin contamination}
\label{sec:spin_cont}

As we saw in the previous sections, the statistical properties of the
model quantum dots obtained from the full SDFT computation are, at
least up to $r_s \simeq 1.3$, well reproduced by the various forms of
the Strutinsky approximation.  However, we also saw that spin
contamination, when present in the SDFT calculations, was actually
degrading the accuracy of the Strutinsky approximation on a case by
case basis.  Indeed, by construction, the Strutinsky approximation as
we presented it cannot involve any spin contamination.  Spin
contamination is a way, in the SDFT calculations, to lower the energy
of the system without changing the total $z$ component of the spin
$S_z$ by having different wavefunctions for the $\al$ and $\be$
orbitals. However, the eigenstates implicit in the Strutinsky approach
are almost identical to those of $\Htf$, $\phi^\s_i$, and the
$\phi^\s_i$ are nearly independent of the spin.

In this section, we shall discuss how this spin contamination
mechanism could be understood within this Strutinsky scheme.  Rather
than trying to consider the most general situation, we will limit
ourselves to the case of even number of particle $N$, and a ground
state $z$ component spin equal to zero.

Let $H_{\rm eff}[\nTF] = -\frac{\hbar^2}{2m} \nabla^2 + \Utf^s(\br)$
be the Thomas-Fermi Hamiltonian defining the orbitals $\phi_i^s$ [see
Eqs.~(\ref{eq:Veff}) and (\ref{eq:schroe})].  What we have done is to
construct a approximate solution of the SDFT equations
Eq.~(\ref{eq:Eks}) as $\tn^\s(\br) = \sum_{j=1}^{N_\s} \phi^\s_j
(\br)$ plus some screening charge. In this respect, an important point
was that the resulting potential change $\delta U^\s$ given by
Eq.~(\ref{eq:deltaUsc}) was such that the matrix element $\langle
\phi_i^s | \delta U^\s |\phi_j^s \rangle$ for $i \neq j$ was
negligible, to first order in $1/g$.

However, finding an approximate solution of the Kohn-Sham equation
implies only that one has an extremum of the spin density functional,
but not necessarily an absolute minimum.  As pointed out earlier,
modifications of the wavefunctions change the electronic density only
if they mix occupied and unoccupied orbitals.  Therefore, when
searching for a new extremum of the spin density functional, with some
chance to actually get the true minimum, a natural choice is to mix
the last occupied orbital $\phi^\s_M$ with the first unoccupied one
$\phi^\s_{M+1}$ (for $M \!=\!N/2 \!=\! N_\al\!=\!N_\be$).

Let us therefore look for approximations $\varphi_i^\s$ to the KS
wavefunctions defined by $\varphi_i^\s = \phi_i^\s$ for $i < M$ and
\begin{eqnarray}
  \varphi^\s_M     & = &  \cos \theta^\s  \phi_M + \sin \theta^\s
  \phi_{M+1} \label{eq:rotation}\\
  \varphi^\s_{M+1} & = & -\sin \theta^\s  \phi_M + \cos \theta^\s
  \phi_{M+1} \; ,
  \nonumber
\end{eqnarray}
with possibly a  different value of the angle  $\theta^\s$ for the two
spins  $\s=\al,\be$.  For these wavefunctions, the Thomas -Fermi
Hamiltonian has a matrix element
\begin{equation}
  \langle \varphi^\s_M | \Htf | \varphi^\s_{M+1} \rangle = \cos \theta^\s
\sin \theta^\s (\et_{M+1} - \et_M)
\end{equation}
in terms of the Thomas-Fermi energies $\et_{M+1}$ and $\et_M$.

This  change in  the wave  functions produces  a modification  of the
densities $\delta  n^\s = |\varphi^\s_M |^2  - \overline{|\varphi|^2}$
which, once screening is taken into account, itself generates a perturbation potential
\begin{equation}
  \delta U^\s (\br) = \sum_{\s'=\al,\be} \int d\br'
  \Vsc^{\s,\s'}(\br,\br') \delta n^{\s'} (\br') \; .
\end{equation}
(The  modification of  the other  wavefunctions, except  for screening
this term, does not play  a role here.)  The self-consistent condition
for the angles $\theta^{(\al,\be)}$ is  that $\langle \varphi^\s_M | \Htf |
\varphi^\s_{M+1}  \rangle  +  \langle  \varphi^\s_M  |\delta  U^\s  (\br)  |
\varphi^\s_{M+1} \rangle = 0$

In order to find matrix elements of $\delta U^\s$, let us consider for
a moment the case $\s=\al$. Eq.~(\ref{eq:rotation}) implies that
\begin{equation}
\varphi^\be_M  = \cos(\theta^\al -\theta^\be)\varphi^\al_M
+ \sin(\theta^\al -\theta^\be)\varphi^\al_{M+1}
\end{equation}
and therefore
\begin{equation}
\langle \varphi_M^\al | \delta U^\al |\varphi_{M+1}^\al \rangle
= 2 \cos(\theta^\al -\theta^\be)\sin(\theta^\al -\theta^\be) I
\end{equation}
with
\begin{eqnarray}
I \! &  = & \!\!\int \!d\br d\br' \varphi_M^\al(\br) \varphi_{M+1}^\al(\br)
\Vsc^{\al,\be}(\br,\br')
\varphi_M^\al(\br') \varphi_{M+1}^\al(\br')
\nonumber \\
 & \simeq &
\langle \hat \Vsc^{\al,\be} \rangle_{\rm fc} /A \;\;
\end{eqnarray}
where the last equality applies in lowest order in $1/g$, that is,
neglecting the fluctuations.

Noting finally that the same reasoning can be applied to the $\be$
orbitals,  we find that to get an  extremum for the spin density
functional the angles $\theta^\al$ and $\theta^\be$ should obey
\begin{eqnarray}
\lefteqn{
\cos\theta^\al \sin\theta^\al (\e_{M+1}-\e_{M}) =} \qquad \\ &  &
 2 \cos(\theta^\al -\theta^\be)\sin(\theta^\al -\theta^\be)
\langle \hat \Vsc^{\al,\be} \rangle_{\rm fc} /A \nonumber 
\end{eqnarray}
and the analogous equation with $ \al$ and $ \be$ interchanged.
Obvious solutions are ($\theta^\al \!=\! \theta^\be \!=\! 0$),
($\theta^\al \!=\! 0,$ $ \theta^\be \!=\! \pi/2$), ($\theta^\al \!=\!
\pi/2,$ $\theta^\be \!=\! 0$), and ($\theta^\al \!=\! \theta^\be \!=\!
\pi/2$).  The first one corresponds to the standard $S \!=\! 0$
(non-spin-contaminated) solution; the other ones involve promotion of
particles from the last occupied orbital to the first unoccupied one,
and so are obviously of higher energy.  However, other solutions may
exist.  Clearly they should fulfill
\begin{equation} 
\cos\theta^\al \sin\theta^\al = \cos\theta^\be
\sin\theta^\be \;;
\end{equation}
that is, up to an irrelevant multiple of $\pi$ phase
\begin{equation}
                 \theta^\be = - \theta^\al \;, \label{eq:extrema1} 
\end{equation}
and therefore
\begin{equation} \label{eq:cont_angle}
\cos2\theta^\al = - \frac{\e_{M+1}-\e_{M}}{4 \langle \hat
  \Vsc^{\al,\be} \rangle_{\rm fc} /A} \; .
\end{equation}

It can be checked that whenever the condition
Eq.~(\ref{eq:cont_angle}) can be fulfilled [ie. when $\e_{M+1}-\e_{M}
\le 4 \langle \hat \Vsc^{\al,\be} \rangle_{\rm fc}$], the
corresponding extremum has an energy smaller by a quantity
$\sin^2(2\theta^\al) [\langle \lambda_s \rangle_{\rm fc} - \langle
\lambda_c \rangle_{\rm fc}] /A$ with respect to the non contaminated
configuration.  In this situation, the spin contaminated $S_z =0$
state will be favored (but its energy still needs to be compared to
the lowest energy state with $S_z =1$).

\section{Discussion}

To summarize our findings, we have seen that up to $r_s$ values of
order one (in practice, 1.3 here), the Strutinsky approximation yields
a ground state total energy with fluctuating part reliable up to
typically five percent of the mean level spacing.  Furthermore, these
errors can be related to the occurrence of spin contamination in the
SDFT calculation, which cannot be reproduced by the Strutinsky scheme
(in its simplest form).  Indeed, as discussed in the last section,
this latter gives, by construction, an approximation to an extremum of
the Kohn-Sham functional for which the $\al$ and $\be$ orbitals are
nearly identical, but not necessarily an approximation to the true
minimum.

Nevertheless, the qualitative properties of the peak spacing and spin
distribution are correctly reproduced by the Strutinsky approximation
(at least up to $r_s$ of order one, for which spin contamination does
not appear to change drastically the distributions).  For a chaotic
confining potential this makes it possible to use the modeling in
terms of Random Matrix Theory (for the energy levels) and Random Plane
Waves (for the eigenstates) introduced in
Ref.~\onlinecite{prb_ullmo_01}. Within this RMT/RPW modeling, and in
the limit of large dots for which fluctuations of the residual
interaction term are small, the main features of the peak spacing and
spin distributions can be understood as arising from the interplay
between one particle level fluctuations and the spin dependence of the
mean residual interaction term [see Eq.~(\ref{eq:DE2_UH})]
\begin{eqnarray} \label{eq:UH_LSDA}
\overline{\Delta E^{(2)}_{\rm SDFT}}(N,S_z) & = &
\lambda^{\rm bare}_s S_z^2 - \frac{\mu_{\rm T}}{2} (\langle \lambda_c
 \rangle_{\rm  fc}  - \langle \lambda_s \rangle_{\rm  fc} )S_z
\nonumber \\ 
 & & \!\!\mbox{} + (\mbox{term depending on $N$ only)} \; . 
\end{eqnarray}

It is useful to compare this expression, as well as the corresponding
distributions based on RMT/RPW modeling such as those in
Fig.~\ref{fig:level4} (which actually take into account the
fluctuations of the residual interaction term), to what is obtained
following the more traditional route to the analysis of peak spacing
and spin distributions for chaotic quantum
dots.\cite{prl_blanter_97,prb_ullmo_01,pr_aleiner_02} In these earlier
approaches, the ground state energy of the quantum dot would, in a way
very similar way to Eq.~(\ref{eq:strut_corr}), be described as the sum
of a large non-fluctuating classical-like term, a one-particle energy
contribution computed for some effective confining potential, and
finally a residual interaction term $\Delta E^{(2)}_{\rm RPA}$.  This
latter would, however, be understood as originating from a weak
interaction $V_{\rm RPA}(\br-\br')$ which in the Random Phase
Approximation can be shown to be just the RPA screened
potential,\cite{prl_blanter_97,pr_aleiner_02} but should be understood
more properly as the residual interaction between quasi-particles in
Landau Fermi liquid spirit.  We shall thus refer to this latter
description as the ``RPA'' approach, although this is slightly
inappropriate.

\begin{figure}[t]
\includegraphics[width=1.72in,clip]{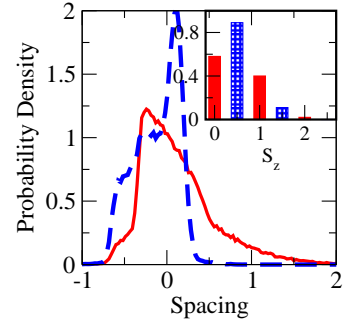}
\caption{(Color online) Same as Fig.~\ref{fig:level4}(b), but with a residual
  interaction modeled by a perturbative calculation in the
  interaction potential $V_{\rm RPA}(\br-\br')$ whose  parameters
  are set by Eqs.~(\ref{eq:FLparam}) and (\ref{eq:zeta}).}
\label{fig:level3}
\end{figure}

\begin{table*}
\begin{tabular}{|c|ccccc|}
\hline
model  & $S=0$ & $S=\frac{1}{2}$ & $S=1$ & $S=\frac{3}{2}$ &  $S=2$ \\
\hline
SDFT   & \ 0.42 $\pm$ 0.03 \  & \    0.76  $\pm$ 0.03   \     & \
0.54 $\pm$ 0.03 \  &   \    0.23 $\pm$ 0.03  \     &  \  0.03 $\pm$
0.01 \ \\
ST$^*$  & 0.34 $\pm$ 0.03 &    0.74   $\pm$ 0.03      & 0.61 $\pm$
0.03 &      
0.25$\pm$ 0.03       &   0.04 $\pm$ 0.01\\
ST$^*$/RPW & 0.28  & 0.68         & 0.62  &      0.30       &   0.09 \\
RPA/RPW (unscreened Cooper) &
          0.48   &  0.84         & 0.49  &      0.16       &   0.03 \\
RPA/RPW (screened Cooper ) &
          0.58   &  0.89    &      0.40 &       0.11       &   0.02 \\
\hline

\end{tabular}
\caption{Ground state spin probability for various model introduced in
  this paper.  The  two   first  lines   correspond
respectively to  the SDFT calculation and [second  form of] Strutinsky
approximation  for  the   quartic  oscillator  systems  introduced  in
section~\ref{sec:KSvsST}.   The statistics are  built from  the ground
state  spin  of  dots containing  100  to  200  electrons, for  a  few
confining  potential corresponding  to an  interaction  parameter $r_s
\simeq 1.3$.  The three last lines are the results of RMT/RPW modeling
for  150 electron  and the  same  value of  the interaction  parameter
$r_s$,  and respectively:  the  interaction derived  from SDFT  (third
line); the  ``RPA''-like interaction using a  screened Cooper channel,
i.e.  such  that  Eqs.~(\ref{eq:FLparam})  and  (\ref{eq:zeta})  apply
(fourth  line); and  the ``RPA''-like  interaction using  a unscreened
Cooper   channel,   i.e.   such   that   Eqs.~(\ref{eq:FLparam})  applies
but $\zeta = J_S$ (fifth line).
} \label{table}
\end{table*}

While the Strutinsky approximation to SDFT gives a residual
interaction which can be understood as a first-order perturbation
(without exchange) in terms of the spin dependent potential
Eq.~(\ref{eq:Vsc}), in contrast one has in the ``RPA'' approach a
residual interaction arising from the perturbative corrections in some
$V_{\rm RPA}(\br,\br')$ (including both direct and exchange) as well
as possibly higher-order terms which turn out to be important for
time-reversal invariant systems (the Cooper series).  Under this
assumption and following exactly the same analysis as the one leading
to Eq.~(\ref{eq:UH_LSDA}), one gets for the mean residual interaction
term in the RPA approach
\begin{eqnarray} \label{eq:UH_RPA}
\overline{\Delta E^{(2)}_{\rm RPA}}(N,S_z) = 
& \! J_S S_z(S_z+1)  +  \zeta (\mu_{\rm T} - 1) S_z
\nonumber \\
& \! \mbox{} + \, (\mbox{term depending on $N$ only)} \;
\end{eqnarray}
(again $\mu_{\rm T} \!=\! 2$ but would be 1 if time-reversal invariance
were  broken). 

In Eq.~(\ref{eq:UH_RPA}), the parameter $J_S$ is equal to $-\langle
\hat V(q) \rangle_{\rm fc}$, where the Fermi circle average is defined
by Eq.~(\ref{eq:mean_fc}) and $\hat V_{\rm RPA}(\bq)$ is the Fourier
transform of $V_{\rm RPA}(\br-\br')$.  More properly however, one
should understand $J_S$ as being related to Fermi liquid parameter
$f^{(a)}_0$ through
\begin{equation} \label{eq:FLparam}
  J_S/\Delta =  f^{(a)}_0 \; .
\end{equation}
In first-order perturbation theory, $\zeta$ would be equal to $J_S$,
but screening associated with higher oder terms in the Cooper channel
somewhat reduces this value.  An analysis following the lines of
Ref.~\onlinecite{prl_ullmo_98} suggests that
\begin{equation} \label{eq:zeta}
\zeta  = \frac{J_s}{1 + \frac{|J_S|}{\Delta} \log(\kf L)} \; ,
\end{equation}
with $\kf$ the Fermi momentum and $L$ the typical size of the
system. We shall assume this in the following discussion, bearing, in
mind that this is true only ``up to logarithmic accuracy''.

The remarkable point here is that $\lambda^{\rm bare}_s(r_s)/\Delta$
is actually the same thing as $f^{(a)}_0 (r_s)$, in the sense that the
Landau Fermi liquid parameter $f^{(a)}_0 (r_s)$ can be interpreted as
the second derivative with respect to the polarization, at fixed total
density, of $\epsilon_{\rm xc}$ the exchange correlation energy per
particle of the uniform electron gas.  This implies that the term
quadratic in $S_z$ of Eqs.~(\ref{eq:UH_LSDA}) and (\ref{eq:UH_RPA})
actually do correspond.

As a consequence if we compare Fig.~\ref{fig:level4}(b), which is
obtained from a RMT/RPW simulation with the interaction corresponding
to the spin density functional, with Fig.~\ref{fig:level3}, obtained
in the same way but with an ``RPA''-like interaction,\footnote{For
this simulation, the fluctuation of the residual interaction term also
depends on the precise way the Cooper channel contribution is
screened.  Fig.~\ref{fig:level4} corresponds to an evaluation of the
variance of these fluctuations for which we have assumed no screening,
and is therefore rather an upper bound.  The fluctuations being in any
case rather small, their magnitude does not change the peak spacing
distribution drastically (their main effect is to broaden the sharp
peak for odd spacings), and a more detailed treatment of the effect of
screening of the Cooper channel on the fluctuations of the residual
interaction term should not change significantly the picture.}  the
differences in spin polarization and in odd/even asymmetry for the
peak spacing distribution can be almost entirely associated with the
different linear $S_z$ terms in Eqs.~(\ref{eq:UH_LSDA}) and
(\ref{eq:UH_RPA}).

To get some further insight into the difference between the two
approaches, let us consider Table~\ref{table} which shows the spin
distributions for an interaction parameter $r_s \simeq 1.3$ and
various approximations discussed in this paper.  Comparing different
pairs of lines gives a sense of the importance of the various issues.
For instance, comparing the second line with the third gives an idea
of how accurate the RMT/RPW model is for the statistical properties of
the real energy levels and eigenfunctions of the quartic oscillator
system, since both lines are based on the same spin dependent
interaction Eq.~(\ref{eq:Vsc}).  The first line compared to the
second, on the other hand, is a measure of how well the Strutinsky
scheme approximates the full SDFT calculation.  Presumably, most of
the difference between these two lines can be associated with the
presence of spin contamination in SDFT -- it is in some sense a
measure of the effectiveness of spin contamination in lowering the
total spin of the system.  Going further down the table, the
difference between the third and the two last lines is a measure of
the impact of different linear terms in $S_z$ in
Eqs.~(\ref{eq:UH_LSDA}) and (\ref{eq:UH_RPA}), without screening the
Cooper channel for the fourth line and with a screened Cooper channel
[according to Eq.~(\ref{eq:zeta})] for the last line of the table.

 From Table~\ref{table}, it appears that within the accuracy of the
RMT/RPW modeling, which seems to be around 5\%, the SDFT result is
compatible with an RPA-like approach if the Cooper channel is not
screened.  In other words, the fact that $ ( \langle \lambda_s
\rangle_{\rm fc} - \langle \lambda_c \rangle_{\rm fc})/2 $ is more
negative that $J_s = \lambda^{\rm bare}_s$ -- producing higher spins
-- is compensated by the effect of spin contamination -- which favors
lower spins.  As seen in Ref.~\onlinecite{hong_rqd}, this compensation
between the two effects seems to exist also for higher values of
$r_s$.  On the other hand, spin contamination is not a sufficiently
strong effect to compensate for the absence of screening of the Cooper
channel.

It remains to decide which of the two approaches is the more
correct. This, in the end, can only be addressed by comparison with
exact calculations for quantum dots (e.g.\ quantum Monte Carlo).  One
argument that may be considered is that in the presence of a
time-reversal breaking term (i.e.  $\mu_{\rm T} = 1$), general
symmetry considerations impose that the mean value of the residual
interaction term is a function of $S(S+1)$, but not independently of
$S^2$ and $S$.  Expressions (\ref{eq:UH_LSDA}) clearly do not fulfill
this constraint, while Eq.~(\ref{eq:UH_RPA}) does.  Since, however,
spin contamination seems to compensate for this difference it might
just be that for time-reversal non-symmetric systems, SDFT and RPA
basically agree.  On the other hand, the screening of the Cooper
channel does not seem to be reproduced by the SDFT calculations, and
this might be the cause of the higher spin found in this approach.

\begin{acknowledgments}

We thank G. Usaj for several valuable conversations and M. Brack for
helpful comments regarding $\hbar$ corrections.  This work was
supported in part by the NSF (DMR-0103003).

\end{acknowledgments}

\end{document}